\documentclass{article}
\usepackage{graphicx}
\RequirePackage{feynarts}
\newcommand{\epe}{\epsilon^\prime/\epsilon}
\newcommand{\be}{\begin{equation}}
\newcommand{\ee}{\end{equation}}
\newcommand{\bea}{\begin{eqnarray}}
\newcommand{\eea}{\end{eqnarray}}
\newcommand{\bra}{\langle}
\newcommand{\ket}{\rangle}

\newcommand{\etal}{{\it et al}.\ }
\begin{document}
\rightline{BNL-HET-03/13}

\begin{center}

{\large\bf Lattice extraction of $ K \rightarrow \pi \pi $
amplitudes to NLO in partially quenched and in full chiral
perturbation theory }

\vspace{.2in}

Jack Laiho$^{*}$\\
 \noindent Department of Physics, Princeton
University, Princeton, NJ\ \ 08544\\
\smallskip
 Amarjit Soni$^{\dag}$ \\
\noindent Theory Group, Brookhaven National Laboratory, Upton, NY\
\
11973\\
\footnotetext{$^*$email: jlaiho@princeton.edu\hskip1.5in
$^\dag$email: soni@bnl.gov}
\end{center}


\begin{quote}
\begin{center}
ABSTRACT
\end{center}

We show that it is possible to construct
$\epsilon^\prime/\epsilon$ to NLO using partially quenched chiral
perturbation theory (PQChPT) from amplitudes that are computable
on the lattice.  We demonstrate that none of the needed amplitudes
require three-momentum on the lattice for either the full theory
or the partially quenched theory; non-degenerate quark masses
suffice. Furthermore, we find that the electro-weak penguin
($\Delta I=3/2$ and 1/2) contributions to
$\epsilon^\prime/\epsilon$ in PQChPT can be determined to NLO
using only degenerate ($m_K=m_\pi$) $K\to\pi$ computations without
momentum insertion.  Issues pertaining to power divergent
contributions, originating from mixing with lower dimensional
operators, are addressed. Direct calculations of $K\to\pi\pi$ at
unphysical kinematics are plagued with enhanced finite volume
effects in the (partially) quenched theory, but in simulations
when the sea quark mass is equal to the up and down quark mass the
enhanced finite volume effects vanish to NLO in PQChPT. In
embedding the QCD penguin left-right operator onto PQChPT an
ambiguity arises, as first emphasized by Golterman and Pallante.
With one version (the ``PQS") of the QCD penguin, the inputs
needed from the lattice for constructing $K\to\pi\pi$ at NLO in
PQChPT coincide with those needed for the full theory. Explicit
expressions for the finite logarithms emerging from our NLO
analysis to the above amplitudes are also given.
\end{quote}

\section{Introduction}

There have been several recent lattice attempts to calculate
$\textrm{Re}(\epe)$, the direct \emph{CP} violating parameter in
$K \to \pi\pi$ decays.  These include attempts with domain wall
fermions by the CP-PACS \cite{noaki} and RBC \cite{blum}
Collaborations.  A notable feature of both of these calculations
is that their central values differ drastically from experiment.
The experiments at CERN \cite{fanti} and Fermilab \cite{alavi}
have yielded an experimental grand average of $\textrm{Re}(\epe) =
(1.8\pm 0.4)\times 10^{-3}$ \cite{pdg}.  The lattice
collaborations find a value $\sim -0.5 \times 10^{-3}$, a
\emph{negative} value, though the groups have made rather severe
approximations.  Such a disagreement between theory and experiment
should not be totally unexpected given the serious approximations
and resulting systematic errors, which have so far been necessary
in order to implement the calculation on the lattice.
\cite{bhatt,pk}

One of these uncontrolled approximations was the use of the
quenched approximation, where the fermion determinant in the path
integral is set to a constant in order to make the problem more
tractable on present day computers.  Another was the use of
leading order chiral perturbation theory (ChPT) to relate
unphysical $K \to \pi$ and $K \to 0$ amplitudes to the physical $K
\to \pi\pi$ amplitudes, as first proposed by \cite{bern}.  Because
of the difficulty of extracting multihadron decay amplitudes from
the lattice, as expressed by the Maiani-Testa theorem
\cite{maiani}, it is much easier to compute the two- and
three-point functions (i.e., $K \to 0$ and $K \to \pi$,
respectively) and use ChPT to extrapolate to the physical matrix
elements.

It is likely that the next-to-leading order (NLO) corrections to
ChPT will be significant for the operators that contribute to
$\textrm{Re}(\epe)$, and should not be neglected.  Unfortunately,
at higher orders in ChPT the number of free parameters that must
be determined from first-principles methods like the lattice
proliferates rapidly.  It has been shown by Cirigliano and
Golowich \cite{cirig} that the dominant electroweak penguin
contributions [(8,8)'s] to $K\to\pi\pi$ can be recovered at NLO
from $K\to\pi$ amplitudes using 4-momentum insertion.  Bijnens,
\etal \cite{bijnens} showed how to obtain most of the low-energy
constants (LEC's) relevant for the case of the (8,1)'s and
(27,1)'s using off-shell $K \to \pi$ Green's functions; not all
LEC's could be determined using this method, though.

In \cite{lin}, it was shown how to obtain physical $K \to \pi\pi$,
$\Delta I =3/2$ [(27,1)'s and (8,8)'s] at NLO from $K \to\pi\pi$
at unphysical (SPQcdR) kinematics accessible to the lattice.  This
method requires 3-momentum insertion, and it is not yet clear if
it can be extended to the $\Delta I =1/2$, $K\to\pi\pi$
amplitudes.  In our previous paper \cite{laiho}, an alternative
method was proposed for constructing the physical $K\to\pi\pi$
amplitudes to NLO for all ($\Delta I=1/2$ and 3/2) operators of
interest.  For the $\Delta I=3/2$ amplitudes this requires $K \to
\overline{K}$; $K\to\pi$, $\Delta I=3/2$; and $K\to\pi\pi$,
$\Delta I=3/2$ at one of (at least) two unphysical kinematics
points where the Maiani-Testa theorem can be bypassed. The two
special kinematics points where this is possible have been
discussed in the literature:  (i) $m^{lat}_K=m^{lat}_\pi$, where
the weak operator inserts energy \cite{berntwo}; and (ii)
$m^{lat}_K=2m^{lat}_\pi$, i.e. at threshold \cite{dawson}.  As in
\cite{laiho}, we refer to these two cases as unphysical kinematics
point 1 (UK1) and point 2 (UK2), respectively.  Finally, it was
also shown in \cite{laiho} how to obtain the physical $K\to\pi\pi$
at NLO for the $\Delta I=1/2$, (8,1) (e.g. $Q_4$ and $Q_6$) and
the mixed $(27,1)\oplus(8,1)$ case (e.g., $Q_2$) using $K \to \pi$
with 4-momentum insertion and $K \to \pi \pi$ at both UK1 and UK2.
Note that the mixed case also requires information obtainable from
the amplitudes needed to get $K\to\pi\pi$, $\Delta I=3/2$ for the
(27,1)'s.  The main purpose of \cite{laiho} was, in fact, to show
that even for the (8,1)'s all of the information needed to
construct $K \to \pi\pi$ to NLO in ChPT could be obtained from
amplitudes that can be computed on the lattice, at least in
principle.

There are other unphysical kinematics values for the $K\to\pi\pi$
amplitudes where the initial and final state mesons are at rest
that bypass the Maiani-Testa theorem. These kinematics are similar
to UK1 in that they require energy insertion, but with $m_K \neq
m_\pi$. We call this set of kinematics UKX.  This corresponds to
the SPQcdR kinematics with both pions at rest \cite{lin}.  Lattice
calculations at these values of the kinematics are important given
that the calculation at UK1 has difficulties \cite{linthree}, and
also given that it will be important to determine the NLO LEC's in
as many ways as possible for additional redundancy. Even if it is
difficult or impossible to obtain the necessary NLO low energy
constants for the (8,1)'s at UK1, one can obtain the same
information using UKX. Thus, all information for the (8,1)'s can
be determined to NLO without using UK1, the difficulties of which
are discussed in Section 8 and in the note added in revision.
Results at UKX are also given in Section 8.

In this work we show that where 4-momentum insertion is required
for any of the amplitudes needed according to the prescription of
\cite{laiho}, it suffices to allow only energy insertion at the
weak operator such that the initial and final state mesons are at
rest.  This means that the $K\to\pi\pi$ amplitudes can be
constructed to NLO using non-degenerate quarks, but without using
3-momentum insertion, making the computation much more economical.

Another approach to $K\to \pi\pi$ and $\epe$ amplitudes has been
proposed by Lellouch and Luscher \cite{lell} in which finite
volume correlation functions on the lattice are used to extract
physical amplitudes without recourse to ChPT, at least in
principle.  This method is expected to be difficult
computationally, but a way of reducing the cost of the
Lellouch-Luscher method has been proposed \cite{christ}.  An
alternative method to obtain $K\to\pi\pi$ amplitudes to all orders
in ChPT has been proposed by \cite{buchler}; this proposal makes
use of dispersion relations.  Both of the above methods depend
crucially on unitarity, so it is unclear if they can be
implemented with partially quenched lattice simulations.

Although NLO ChPT may not be the final answer, it is more reliable
than leading order, and it is useful to have the NLO expressions
even to extract the leading order LEC's from the lattice data.
Since the lattice data that will be generated in the near term
will be in the (partially) quenched approximation, it is necessary
to have the corresponding amplitudes in partially quenched ChPT.
Therefore, in this paper, we present the partially quenched
expressions for the quantities of greatest interest for
$\textrm{Re}(\epe)$, namely the amplitudes for the (8,1) and (8,8)
operators. For the partially quenched amplitudes we assume that
all relevant quark masses are small compared to the $\eta'$ mass,
so that the $\eta'$ can be integrated out, and the LEC's of the
partially quenched theory coincide with those of the full theory
when the number of sea quarks is three \cite{sharpe}.

For the $\Delta I =1/2$ amplitudes there is an additional
complication involving eye diagrams having to do with the sum over
quarks in the penguin operators \cite{goltthree}. For the
left-right gluonic penguin operators the two possible choices
correspond to what we will call the PQS (partially quenched
singlet) method and the PQN (partially quenched non-singlet)
method. They are discussed in detail in Section 6.1. It is
important to note that only for the PQS method can the LEC's
sufficient to construct $\epe$ to NLO be determined, whereas it is
not clear if the PQN method can be extended to NLO.  Indeed, a
significant advantage of the PQS implementation is that the
ingredients needed from the lattice to obtain all $K\to\pi\pi$
amplitudes to NLO in PQChPT are the same as in the full theory.
Therefore, the PQS method is used to compute the NLO amplitudes in
this paper. Finally, it should be mentioned that the $\Delta
I=1/2$, $K\to\pi\pi$ amplitudes receive enhanced finite volume
contributions in the partially quenched theory
\cite{bernfour,colang}. However, when $m_{sea}=m_u=m_d$, the
infra-red divergences in the $K\to\pi\pi$ amplitudes (at UK1 and
UK2) vanish in PQChPT in the infinite volume Minkowski space
amplitudes.  In an earlier version of this paper we had pointed
out that it would be important to study the finite volume effects
of these amplitudes; the corresponding finite volume Euclidian
Green's functions were calculated by \cite{linthree} while this
work was in revision. \footnote{Ref \cite{linthree} found that the
infra-red problems do not vanish for UK1 finite volume Euclidean
correlation functions in the partially quenched theory; for
further details, see our note added in revision.}

In the partially quenched theory, it is possible to construct the
(8,8) $K\to\pi\pi$ amplitudes to NLO using only degenerate valence
quark masses in $K\to\pi$, along with $K\to0$ in order to perform
the power divergent subtraction in the $\Delta I=1/2$ case.
Additional redundancy is possible if one uses nondegenerate
valence quark masses in the $K\to\pi$ calculation.

The content of the paper is as follows.  Section 2 briefly reviews
the formalism of effective four-fermion operators in a standard
model calculation.  Section 3 reviews ChPT and the realization of
the effective four-quark operators in terms of ChPT operators for
weak processes.  Section 4 reviews partially quenched chiral
perturbation theory and how it can be extended to the electroweak
sector.  Section 5 presents results for the full theory,
demonstrating that for all the amplitudes considered in
\cite{laiho}, 3-momentum insertion is not essential and
non-degenerate quark masses suffices to construct $K \to \pi\pi$
to NLO. In Section 6 a discussion of the treatment of eye-diagrams
in the partially quenched theory is given, as well as a comparison
of PQS and PQN results at leading order according to the papers by
Golterman and Pallante \cite{goltthree}.  Sections 7 and 8 present
the main results of this paper, showing how to obtain the $K \to
\pi \pi$ amplitudes needed for $\textrm{Re}(\epe)$ in the
partially quenched theory from quantities which can be computed
directly on the lattice. Section 7 deals with the (8,8)
amplitudes, while Section 8 deals with the (8,1)'s.  Section 9
discusses the checks done on the various one-loop logarithmic
expressions. Section 10 presents the conclusion. Section 11 is a
note added in revision. The finite logarithm contributions to the
relevant amplitudes are presented in a set of Appendixes.  Errors
in Eqs (31, D6) of \cite{laiho} are corrected in Appendix F.

\section{Effective Four Quark Operators}

In the Standard Model, the nonleptonic interactions can be
expressed in terms of an effective $ \Delta S=1 $ hamiltonian
using the operator product expansion \cite{ciuch,bucha},

\begin{equation}\label{1}
    \langle \pi \pi |{\cal H}_{\Delta S=1}|K\rangle =
    \frac{G_{F}}{\sqrt{2}} \sum V_{CKM}^{i} c_{i}(\mu)
    \langle \pi \pi|Q_{i}|K\rangle_{\mu},
\end{equation}

\noindent where $V_{CKM}^{i}$ are the relevant combinations of CKM
matrix elements, $ c_{i}(\mu) $ are the Wilson coefficients
containing the short distance perturbative physics, and the matrix
elements $ \langle \pi \pi|Q_{i}|K\rangle_{\mu} $ must be
calculated nonperturbatively. The four quark operators are

\begin{equation}\label{2}
    Q_{1}=\overline{s}_{a} \gamma_{\mu} (1-\gamma^{5}) d_{a}
    \overline{u}_{b}\gamma^{\mu} (1-\gamma^{5}) u_{b},
\end{equation}
\begin{equation}
Q_{2}=\overline{s}_{a} \gamma_{\mu} (1-\gamma^{5}) d_{b}
    \overline{u}_{b}\gamma^{\mu} (1-\gamma^{5}) u_{a},
\end{equation}
\begin{equation}
Q_{3}=\overline{s}_{a} \gamma_{\mu} (1-\gamma^{5}) d_{a} \sum_{q}
    \overline{q}_{b}\gamma^{\mu} (1-\gamma^{5}) q_{b},
    \end{equation}
    \begin{equation}
Q_{4}=\overline{s}_{a} \gamma_{\mu} (1-\gamma^{5}) d_{b} \sum_{q}
    \overline{q}_{b}\gamma^{\mu} (1-\gamma^{5}) q_{a},
    \end{equation}
     \begin{equation}
Q_{5}=\overline{s}_{a} \gamma_{\mu} (1-\gamma^{5}) d_{a} \sum_{q}
    \overline{q}_{b}\gamma^{\mu} (1+\gamma^{5}) q_{b},
     \end{equation}
     \begin{equation}
Q_{6}=\overline{s}_{a} \gamma_{\mu} (1-\gamma^{5}) d_{b} \sum_{q}
    \overline{q}_{b}\gamma^{\mu} (1+\gamma^{5}) q_{a},
    \end{equation}
    \begin{equation}
Q_{7}=\frac{3}{2} \overline{s}_{a} \gamma_{\mu} (1-\gamma^{5})
d_{a}
     \sum_{q} e_{q}
    \overline{q}_{b}\gamma^{\mu} (1+\gamma^{5}) q_{b},
    \end{equation}
    \begin{equation}
Q_{8}=\frac{3}{2} \overline{s}_{a} \gamma_{\mu} (1-\gamma^{5})
d_{b}
     \sum_{q} e_{q}
    \overline{q}_{b}\gamma^{\mu} (1+\gamma^{5}) q_{a},
     \end{equation}
      \begin{equation}
Q_{9}=\frac{3}{2} \overline{s}_{a} \gamma_{\mu} (1-\gamma^{5})
d_{a}
     \sum_{q} e_{q}
    \overline{q}_{b}\gamma^{\mu} (1-\gamma^{5}) q_{b},
    \end{equation}
    \begin{equation}
Q_{10}=\frac{3}{2} \overline{s}_{a} \gamma_{\mu} (1-\gamma^{5})
d_{b}
     \sum_{q} e_{q}
    \overline{q}_{b}\gamma^{\mu} (1-\gamma^{5}) q_{a}.
\end{equation}

In the effective theory $ Q_{1} $ and $ Q_{2} $ are the
current-current weak operators, $ Q_{3}-Q_{6} $ are the operators
arising from QCD penguin diagrams, while $ Q_{7}-Q_{10} $ are the
operators arising from electroweak penguin diagrams.  Note that
the definitions of $Q_1$ and $Q_2$ are different from our previous
paper \cite{laiho}.  After a Fierz transformation, one can see
that the definitions of the two operators are switched.  We have
changed the definitions of $Q_1$ and $Q_2$ to be consistent with
the basis used by RBC \cite{blum} and that of \cite{ciuch}; this
does not, of course, effect any of the results of our previous
paper.

\section{Chiral Perturbation Theory}

    Chiral perturbation theory (ChPT) is an effective quantum
    field theory where the quark and gluon degrees of freedom have
    been integrated out, and is expressed only in terms of the
    lowest mass pseudoscalar mesons \cite{georgi}.  It is a perturbative
    expansion about small quark masses and small momentum of the low mass
    pseudoscalars.  The effective Lagrangian is made up of
    complicated nonlinear functions of the pseudoscalar fields,
    and is nonrenormalizable, making it necessary to introduce
    arbitrary constants at each order in perturbation theory.  In
    such an expansion, operators of higher order in the momentum
    (terms with increasing numbers of derivatives) or mass appear at
    higher order in the perturbative expansion.  The most general
    set of operators at a given order can be constructed out of
    the unitary chiral matrix field $ \Sigma $, given by

    \begin{equation}\label{3}
    \Sigma = \exp \left[\frac {2i\phi^{a}\lambda^{a}}{f}\right],
\end{equation}

    \noindent where $ \lambda^{a} $ are proportional to the Gell-Mann
    matrices with $\textrm{tr}(\lambda_a\lambda_b)=\delta_{ab}$, $ \phi^{a} $ are the
real pseudoscalar-meson fields,
    and $ f $ is the meson decay constant in the chiral limit, with $ f_{\pi} $
    equal to 130 MeV in our convention.

    At leading order $ [O(p^{2})] $ in ChPT, the strong Lagrangian
    is given by

    \begin{equation}\label{4}
    {\cal L}^{(2)}_{st}=\frac{f^{2}}{8}
    \textrm{tr}[\partial_{\mu}\Sigma\partial^{\mu}\Sigma] +
    \frac{f^{2}B_{0}}{4}\textrm{tr}[\chi^{\dag}\Sigma+\Sigma^{\dag}\chi],
 \end{equation}

\noindent where $ \chi= (m_u,m_d,m_s)_{\rm diag} $ and

 $ B_{0}= \frac{m^{2}_{\pi^{+}}}{m_{u}+m_{d}}=
\frac{m^{2}_{K^{+}}}{m_{u}+m_{s}}=\frac{m^{2}_{K^{0}}}{m_{d}+m_{s}}.\\
$

\noindent The leading order weak chiral Lagrangian is given by
\cite{bern,cirig}

\begin{eqnarray}\label{5}
    {\cal L}^{(2)}_{W}&  =  &\alpha_{88}\textrm{tr}[\lambda_{6}\Sigma Q \Sigma^{\dag}]
               +
    \alpha_{1}\textrm{tr}[\lambda_{6}\partial_{\mu}\Sigma\partial^{\mu}\Sigma^{\dag}]
    +\alpha_{2} 2B_{0}
    \textrm{tr}[\lambda_{6}(\chi^{\dag}\Sigma+\Sigma^{\dag}\chi)] \nonumber \\
    \!\!& & +
    \alpha_{27}t^{ij}_{kl}(\Sigma\partial_{\mu}\Sigma^{\dag})^{k}_{i}
    (\Sigma\partial^{\mu}\Sigma^{\dag})^{l}_{j} + \textrm{H.c.},
\end{eqnarray}

\noindent where $t^{ij}_{kl}$ is symmetric in $i, j$ and $k, l$,
traceless on any pair of upper and lower indices with nonzero
elements $t^{13}_{12}=1$, $t^{23}_{22}=1/2$ and
$t^{33}_{32}=-3/2$. Also, $ Q $ is the quark charge matrix, $
Q=1/3(2,-1,-1)_{\rm diag}$
 and $ (\lambda_6)_{ij}= \delta_{i3}\delta_{j2}$.  The reason
 $\lambda_6$ enters these expressions is because it picks out the
 $s$ to $d$, $\Delta S=1$ transition.

The terms in the weak Lagrangian can be classified according to
their chiral transformation properties under
$\textrm{SU}(3)_{L}\times \textrm{SU}(3)_{R}$.  The first term in
(14) transforms as $ 8_{L}\times 8_{R} $ under chiral rotations
and corresponds to the electroweak penguin operators $ Q_{7} $ and
$ Q_{8} $.  The next two terms in (14) transform as $ 8_{L}\times
1_{R} $, while the last transforms as $ 27_{L}\times 1_{R} $ under
chiral rotations.  All ten of the four quark operators of the
effective weak Lagrangian have a realization in the chiral
Lagrangian differing only in their transformation properties and
the values of the low energy constants which contain the
non-perturbative dynamics of the theory.

For the transition of interest, $ K \rightarrow \pi \pi $, the
operators can induce a change in isospin of $\frac12$ or $\frac32$
leading to a final isospin state of the pions of 0 or 2,
respectively. We can then classify the isospin components of the
four quark operators according to their
transformation properties \cite{noaki,blum}:\\

\hspace{1cm}\noindent $ Q^{1/2}_{1}, Q^{1/2}_{2}, Q^{1/2}_{9},
Q^{1/2}_{10}:
8_{L}\times 1_{R} \oplus 27_{L}\times 1_{R}$;\\

\hspace{1cm}\noindent $Q^{3/2}_{1}, Q^{3/2}_{2}, Q^{3/2}_{9},
Q^{3/2}_{10}:  27_{L}\times 1_{R}$;\\

\hspace{2cm}\noindent $Q^{1/2}_{3}, Q^{1/2}_{4}, Q^{1/2}_{5},
Q^{1/2}_{6}: 8_{L}\times
1_{R}$;\\

\hspace{2cm}\noindent$Q^{1/2}_{7}, Q^{1/2}_{8}, Q^{3/2}_{7},
Q^{3/2}_{8}: 8_{L}\times
8_{R}.\\\\
$

Note that $ Q_{3}-Q_{6} $ are pure isospin $\frac12$ operators.
At NLO the strong Lagrangian involves 12 additional operators with
undetermined coefficients.  These were introduced by Gasser and
Leutwyler in \cite{gass}.  The complete basis of counterterm
operators for the weak interactions with $ \Delta S=1, 2 $ was
treated by Kambor, Missimer and Wyler in \cite{kambor} and
\cite{ecker}. A minimal set of counterterm operators contributing
to $ K \rightarrow \pi $ and $ K \rightarrow \pi \pi $ for the $
(8_{L},1_{R})$ and $ (27_{L},1_{R})$ cases is given by
\cite{golt}, with the effective Lagrangian

\begin{equation}\label{7}
    {\cal L}^{(NLO)}_{W}= \sum e_{i} {\cal O}^{(8,1)}_{i}+ \sum
    d_{i}{\cal O}^{(27,1)}_{i}+ \sum
    c_{i}{\cal O}^{(8,8)}_{i},
\end{equation}

\begin{equation}
\begin{array}{ll}
 {\cal O}^{(8,1)}_{1}= \textrm{tr}[\lambda_{6} S^2], & {\cal O}^{(27,1)}_{1}=t^{ij}_{kl}
(S)^{k}_{i}(S)^{l}_{j},\\
{\cal O}^{(8,1)}_{2}= \textrm{tr}[\lambda_{6} S] \textrm{tr}[S], &
{\cal O}^{(27,1)}_{2}=t^{ij}_{kl}
(P)^{k}_{i}(P)^{l}_{j},\\
{\cal O}^{(8,1)}_{3}=\textrm{tr}[\lambda_{6} P^{2}], &
{\cal O}^{(27,1)}_{4}=t^{ij}_{kl}(L_{\mu})^{k}_{i}(\{L^{\mu},S\})^{l}_{j}, \\
{\cal O}^{(8,1)}_{4}=\textrm{tr}[\lambda_{6} P] \textrm{tr}[P], &
{\cal O}^{(27,1)}_{5}=t^{ij}_{kl}
(L_{\mu})^{k}_{i}([L^{\mu},P])^{l}_{j},\\
{\cal O}^{(8,1)}_{5}=\textrm{tr}[\lambda_{6}[S,P]], & {\cal
O}^{(27,1)}_{6}=t^{ij}_{kl}
(S)^{k}_{i}(L^{2})^{l}_{j},\\
{\cal O}^{(8,1)}_{10}=\textrm{tr}[\lambda_{6} \{S,L^{2}\}], &
{\cal O}^{(27,1)}_{7}=t^{ij}_{kl}
(L_{\mu})^{k}_{i}(L^{\mu})^{l}_{j}
\textrm{tr}[S],\\
{\cal O}^{(8,1)}_{11}=\textrm{tr}[\lambda_{6} L_{\mu} S L^{\mu}],
& {\cal O}^{(27,1)}_{20}=t^{ij}_{kl}
(L_{\mu})^{k}_{i}(\partial_{\nu}W^{\mu
\nu})^{l}_{j},\\
{\cal O}^{(8,1)}_{12}=\textrm{tr}[\lambda_{6} L_{\mu}]
\textrm{tr}[\{L^{\mu},S\}], & {\cal O}^{(27,1)}_{24}=t^{ij}_{kl}
(W_{\mu \nu})^{k}_{i}(W^{\mu
 \nu})^{l}_{j},\\
{\cal
O}^{(8,1)}_{13}=\textrm{tr}[\lambda_{6} S] [L^{2}],\\
{\cal
O}^{(8,1)}_{15}=\textrm{tr}[\lambda_{6} [P,L^{2}]],\\
{\cal
O}^{(8,1)}_{35}=\textrm{tr}[\lambda_{6}\{L_{\mu},\partial_{\nu}W^{\mu
\nu}\}], \qquad\qquad\qquad\\
\bigskip
\bigskip
{\cal O}^{(8,1)}_{39}=\textrm{tr}[\lambda_{6} W_{\mu \nu} W^{\mu
\nu}], \\
 {\cal O}^{(8,8)}_{1}= \textrm{tr}[\lambda_{6} L_{\mu}\Sigma^{\dag} Q \Sigma L^{\mu}],\\
 {\cal O}^{(8,8)}_{2}= \textrm{tr}[\lambda_{6} L_{\mu}]\textrm{tr}[\Sigma^{\dag} Q \Sigma L^{\mu}],\\
 {\cal O}^{(8,8)}_{3}= \textrm{tr}[\lambda_{6} \{\Sigma^{\dag} Q \Sigma, L^{2}\}],\\
 {\cal O}^{(8,8)}_{4}= \textrm{tr}[\lambda_{6} \{\Sigma^{\dag} Q \Sigma, S\}],\\
 {\cal O}^{(8,8)}_{5}= \textrm{tr}[\lambda_{6} [\Sigma^{\dag} Q \Sigma, P]],\\
 {\cal O}^{(8,8)}_{6}= \textrm{tr}[\lambda_{6} \Sigma^{\dag} Q \Sigma] \textrm{tr}[S],
\end{array}
\end{equation}
\medskip

\noindent with $ S=2B_{0}(\chi^{\dag}\Sigma + \Sigma^{\dag}\chi$),
$P=2B_{0}(\chi^{\dag}\Sigma-\Sigma^{\dag}\chi$), $L_{\mu}=i
\Sigma^{\dag}\partial_{\mu}\Sigma$ , and $W^{\mu
\nu}=2(\partial_{\mu}L_{\nu}+\partial_{\nu}L_{\mu})$.

This list is identical to that of Bijnens et al. \cite{bijnens}
for the $(27,1)$'s and the $(8,1)$'s, except for the inclusion of
$ {\cal O}^{(8,1)}_{35,39} $ and $ {\cal O}^{(27,1)}_{20,24} $
which contain surface terms, and so cannot be absorbed into the
other constants for processes which do not conserve 4-momentum at
the weak vertex.  Since we must use 4-momentum insertion in a
number of our amplitudes, these counterterms must be considered,
and they are left explicit even in the physical amplitudes.  The
list of $(8,8)$ operators is that of Cirigliano and Golowich
\cite{cirig}.

The divergences associated with the counterterms have been
obtained in \cite{cirig}, \cite{bijnens}, and \cite{kambor}. The
subtraction procedure can be defined as

\begin{equation}\label{8}
    e_{i}=e^{r}_{i}+\frac{1}{16
    \pi^{2}f^{2}}\left[\frac{1}{d-4}+\frac{1}{2}(\gamma_{E}-1-\ln 4
    \pi)\right]2(\alpha_{1}\varepsilon_{i}+\alpha_{2}\varepsilon'_{i}),
\end{equation}

\begin{equation}\label{9}
    d_{i}=d^{r}_{i}+\frac{1}{16
    \pi^{2}f^{2}}\left[\frac{1}{d-4}+\frac{1}{2}(\gamma_{E}-1-\ln 4
    \pi)\right]2 \alpha_{27} \gamma_{i},
\end{equation}

\begin{equation}\label{9}
    c_{i}=c^{r}_{i}+\frac{1}{16
    \pi^{2}f^{2}}\left[\frac{1}{d-4}+\frac{1}{2}(\gamma_{E}-1-\ln 4
    \pi)\right]2 \alpha_{88} \eta_{i},
\end{equation}

\noindent with the divergent pieces, $ \varepsilon_{i},
\varepsilon'_{i}, \gamma_{i} $, $\eta_{i}$ given in Table 1.

It is also necessary for the method of this paper to consider the
$O(p^{4})$ strong Lagrangian, which was first given by Gasser and
Leutwyler, ${\cal L}^{(4)}_{st}=\sum L_{i}{\cal O}^{(st)}_{i}$.

\begin{table}[htbp]
\caption{The divergences in the weak $O(p^4)$ counterterms,
$e_{i}$'s and $d_{i}$'s, for the (8,1)'s and (27,1)'s,
respectively, and the divergences in the weak $O(p^2)$
counterterms, the $c_{i}$'s for the (8,8)'s. \label{tabone}}
\begin{center}
\begin{tabular}{|c|c|c|c|c|c|c|}
  \hline
   $ e_{i}$ & $ \varepsilon_{i}$ & $ \varepsilon'_{i} $ & $ d_{i} $ & $
   \gamma_{i} $ & $ c_{i} $ & $ \eta_{i} $ \\
  \hline
  1 & $ 1/4 $ & $ 5/6 $ &  1 & $ -1/6 $ & 1 & 0 \\
  2 & $ -13/18 $ & $ 11/18 $ &   2 & 0  & 2 & $-2$ \\
  3 & $ 5/12 $ & 0 &   4 & 3  & 3 & $-3/2$ \\
  4 & $ -5/36 $ & 0 &   5 & 1 & 4 & 3/2 \\
  5 & 0 & $ 5/12 $ &  6 & $ -3/2 $ & 5 & 0 \\
  10 & $ 19/24 $ & $ 3/4 $ &   7 & 1  & 6 & 1 \\
  11 & $ 3/4 $ & 0 &  20 & $ 1/2 $ &&\\
  12 & $ 1/8 $ & 0 &  24 & $ 1/8 $ &&\\
  13 & $ -7/8 $ & $ 1/2 $ & &  &&\\
  15 & $ 23/24 $ & $ -3/4 $ &  & &&\\
  35 & $ -3/8 $ & 0 & & &&\\
  39 & $ -3/16 $ & 0 & &  &&\\
  \hline
\end{tabular}
\end{center}
\end{table}

\noindent The strong $O(p^{4})$ operators relevant for
this calculation are the following \cite{gass}:\\

  \noindent \be \begin{array}{ll}{\cal
O}^{(st)}_{1}=\textrm{tr}[L^{2}]^{2},\\
    {\cal
O}^{(st)}_{2}=\textrm{tr}[L_{\mu}L_{\nu}]\textrm{tr}[L^{\mu}L^{\nu}],\\
     {\cal
O}^{(st)}_{3}=\textrm{tr}[L^{2}L^{2}],\\
     {\cal
O}^{(st)}_{4}=\textrm{tr}[L^{2}]\textrm{tr}[S],\\
     {\cal
O}^{(st)}_{5}=\textrm{tr}[L^{2}S],\\
     {\cal
O}^{(st)}_{6}=\textrm{tr}[S]^{2},\\
    {\cal
O}^{(st)}_{8}=\frac{1}{2} \textrm{tr}[S^{2}-P^{2}].\end{array} \ee\\

    The Gasser-Leutwyler counterterms also contribute to the
    cancellation of divergences in the expressions relevant to
    this paper.  The subtraction is defined similarly to that of
    the weak counterterms,

\begin{equation}\label{10}
    L_{i}=L^{r}_{i}+\frac{1}{16
    \pi^{2}}\left[\frac{1}{d-4}+\frac{1}{2}(\gamma_{E}-1-\ln 4
    \pi)\right] \Gamma_{i},
\end{equation}

\noindent with the divergent parts of the counterterm coefficients
given in Table 2  \cite{gass}.\\

\begin{table}[htbp]
\caption{The divergences in the strong $O(p^{4})$ counterterms,
$\Gamma_{i}$\cite{gass}. \label{tabtwo}}
\begin{center}
\begin{tabular}{|c|c|}
  \hline
  i & $\Gamma_{i}$ \\
  \hline
  1 & 3/32 \\
  2 & 3/16 \\
  3 & 0 \\
  4 & 1/8 \\
  5 & 3/8 \\
  6 & 11/144 \\
  8 & 5/48 \\
  \hline
\end{tabular}
\end{center}
\end{table}

\section{Partially Quenched Chiral Perturbation Theory}

There are two approaches to (partially) quenched QCD, the
supersymmetric formulation \cite{bernthree} and the replica method
\cite{damgaard}. Damgaard and Splittorff claim that the two
methods are equivalent in the context of perturbation theory in
the strong sector.  We choose to follow the original method of
Bernard and Golterman \cite{bernthree} for partially quenched
chiral perturbation theory (PQChPT).  In this method, the valence
quarks are quenched by introducing ``ghost" quarks which have the
same mass and quantum numbers as the valence quarks but opposite
statistics. As in \cite{golt}, we consider a theory with $n$
quarks and $N$ sea quarks, so that there are $n-N$ valence and
$n-N$ ghost quarks. The valence quarks have arbitrary mass, while
the sea quarks are all degenerate.  The symmetry group of the
action is $ SU(n|n-N)_{L} \otimes SU(n|n-N)_{R}$.

In the partially quenched case, the chiral field

\begin{equation}
\Sigma = \exp \left[\frac {2i\phi^{a}\lambda^{a}}{f}\right],
\end{equation}

\noindent has $\phi^{a}\lambda^{a}$ replaced by a
$(2n-N)\times(2n-N)$ matrix,

\begin{equation}
\Phi \equiv \left(%
\begin{array}{cc}
  \phi & \chi^{\dag} \\
  \chi & \widetilde{\phi} \\
\end{array}%
\right),
\end{equation}

\noindent where $\phi$ is an $n \times n$ matrix containing the
pseudoscalar meson fields comprised of normal valence and sea
quarks. $\widetilde{\phi}$ is an $(n-N) \times (n-N)$ matrix
comprised of ghost-antighost quarks, while $\chi^\dag$ is an $n
\times (n-N)$ matrix of Goldstone fermions comprised of quarks and
anti-ghosts. The most general set of operators can be constructed
out of $\Sigma$, and these operators can be written in block form
as

\begin{equation}
U = \left(%
\begin{array}{cc}
  A & B \\
  C & D \\
\end{array}%
\right),
\end{equation}

\noindent where the sub-matrices have the same dimension as the
elements of $\Phi$, above.  The transition to the partially
quenched theory is made by replacing $\phi^{a}\lambda^{a}$ by the
above $(2n-N) \times (2n-N)$ matrix, $\Phi$, and replacing the
traces in the operators with supertraces, defined as

\begin{equation}
\textrm{str}(U)=\textrm{tr}(A)-\textrm{tr}(D).
\end{equation}

As a practical matter, in almost all of the NLO diagram
calculations considered in this paper, the minus sign in the
supertrace is cancelled by an additional minus sign coming from
anticommuting pseudo-fermion fields.  The bare mass of a
pseudoscalar meson is given by

\begin{equation}
m_{ij}^2 = B_{0}(m_i + m_j),
\end{equation}

\noindent where $m_i$ and $m_j$ are the masses of the two quarks
that form the meson.  We define $m_{33}$ to be the tree-level
meson mass of two valence strange quarks, as in \cite{golt}

\begin{equation}
m^2_{33}=2m^2_K-m^2_\pi.
\end{equation}

The tree-level mass of a meson made from the \emph{i}th valence
quark and a sea quark is

\bea
m^2_{iS}=B_0 (m_i+m_S)=\frac12(m^2_{ii}+m^2_{SS}), \nonumber \\
   i=u,d,s.
\eea

In this paper we consider only the case where the $\eta'$ has been
integrated out.  Thus, the results are applicable to lattice
calculations only when both sea and valence quark masses are small
compared to $m_{\eta'}$.  Although it may be difficult
computationally, this is precisely the case in which the LEC's of
PQChPT are the same as those of full QCD when the number of sea
quarks is three \cite{sharpe}.  This is because the LEC's are
independent of quark mass even if one varies sea and valence
masses separately.  In order that the LEC's of PQChPT be those of
the real world, the sea and valence quarks must be small enough
that the $\eta'$ decouples, and its effects are integrated out the
same way in both PQChPT and in full ChPT.

The Minkowski space propagators for the flavor diagonal elements
of $\Phi$ are given by analytically continuing the Euclidean
expression in \cite{golt}

\begin{equation}
\Delta_{ij} =
\frac{\delta_{ij}\epsilon_{i}}{p^{2}-m_{ii}^2+i\varepsilon}
-\frac{1}{N}\left(\frac{1}{p^{2}-m_{ii}^2+i\varepsilon}
+\frac{m_{jj}^2-m_{SS}^2}{(p^2-m_{ii}^2+i\varepsilon)(p^2-m_{jj}^2+i\varepsilon)}\right),
\end{equation}

\noindent where

\begin{equation}
\epsilon_{i}=
\left\{%
\begin{array}{ll}
    +1, & \hbox{for $1 \leq i \leq n$   (valence and sea);} \\
    -1, & \hbox{for $n+1 \leq i \leq 2n-N$   (ghost).} \\
\end{array}%
\right.
\end{equation}

At LO (NLO), the operators in PQChPT are still given by Eqs.
(13,14) [Eqs. (16,20)], but with tr $\rightarrow$ str for all
operators. In the extension to the partially quenched case,

\begin{equation}
\lambda_6 \rightarrow \left(%
\begin{array}{cc}
  \lambda_6 & 0 \\
  0 & 0 \\
\end{array}%
\right),
\end{equation}
\medskip

\noindent in block diagonal form, and the mass matrix,

\begin{equation}
\chi \rightarrow \textrm{diag}(m_u, m_d, m_s, m_{sea},..., m_u,
m_d, m_s).
\end{equation}

There is a choice in how to embed the quark charge matrix in the
partially quenched theory, and this will affect the $\Delta
I=1/2$, (8,8) amplitudes considered in this paper.  If one wants
to partially quench the electroweak penguin operators, then the
ghost quark charges should be the same as the corresponding
valence quark charges. If, on the other hand, one wants to allow
valence quarks to couple to photons and Z's, then the ghost quark
charges should be set to zero so they do not appear in, and
therefore cancel, the electroweak valence quark loops.  We present
amplitudes in this paper for both choices. Also, since we choose
the sea quarks to have degenerate mass, the sum of the sea quark
charges is the only quantity involving the sea quark charge that
contributes. This is zero for three flavors, and in this paper we
keep this true for arbitrary sea quark number, $N$, by setting the
sea quark charge to zero.

Also in the partially quenched case, the coefficient of the
counterterm divergence depends on the number of sea quarks, $N$
\cite{colang}. The $N$ dependence of the necessary coefficients
for the $(8,1)$'s was calculated following \cite{kambor,kamb}, and
the results are presented in Table 3.  This paper uses a different
basis from \cite{colang} for the $(8,1)$'s, and also several more
LEC's appear here, so the calculation was redone for this work.
The usual method was employed, expanding the action around the
classical solution (background field method) and using a heat
kernel expansion.  The $N$ dependence of the coefficients of the
divergent parts of the $(8,8)$ counterterms was given in
\cite{cirig}. These values are also presented in Table 3.

\begin{table}[htbp]
\caption{The $N$ dependence of the divergences in the NLO
counterterms, $e_i$'s and $c_i$'s for the (8,1)'s and (8,8)'s,
respectively. \label{tabthree}}
\begin{center}
\begin{tabular}{|c|c|c|c|c|}
  \hline
  $e_i$ & $\varepsilon_i$ & $\varepsilon_i'$ & $c_i$ & $\eta_i$ \\
  \hline
  1 & $-N/4+3/N$ & $N/2-2/N$ & 1 & 0 \\
  2 & $-1/2-2/N^2$ & $1/2+1/N^2$ & 2 & $-2$ \\
  3 & $N/4-1/N$ & 0 & 3 & $-N/2$ \\
  5 & 0 & $N/4-1/N$ & 4 & $N/2$ \\
  10 & $N/8+1/(2N)$ & $N/4$ & 5 & 0 \\
  11 & $N/2-3/N$ & 0 & 6 & 1 \\
  13 & $-3/4$ & 1/2 &  &  \\
  14 & 1/4 & 0 &  &  \\
  15 & $3N/8-1/(2N)$ & $-N/4$ &  &  \\
  35 & $-N/8$ & 0 &  &  \\
  39 & $-N/16$ & 0 &  &  \\
  \hline
\end{tabular}
\end{center}
\end{table}

It is necessary to include an additional (8,1) operator, ${\cal
O}^{(8,1)}_{14}= \textrm{str}[\lambda_{6} L^2]\textrm{str}[S]$, in
this analysis of the partially quenched case since it can no
longer be written as a linear combination of the other operators
via the Cayley-Hamilton theorem.  In the case of full ChPT, the
operator ${\cal O}^{(8,1)}_{14}$ is absorbed into the other
operators, ${\cal O}^{(8,1)}_{10}$, ${\cal O}^{(8,1)}_{11}$,
${\cal O}^{(8,1)}_{12}$ and ${\cal O}^{(8,1)}_{13}$.  Since
$e_{14}$ has a divergent part, the coefficients of the divergences
of the other four operators are modified (for $N=3$) from the
values in Table 1.  Note that $e_4$ and $e_{12}$ have been omitted
in Table 3.  These LEC's do not appear in any of the amplitudes of
interest in this paper.

The Gasser-Leutwyler counterterms also contribute to the
cancellation of divergences in this paper in the partially
quenched case.  The $N$ dependence of the coefficients,
$\Gamma_i$, is given in Table 4.

\begin{table}[htbp]
\caption{$N$ dependence of the divergences in the strong
$O(p^{4})$ counterterms, $\Gamma_{i}$\cite{colang}.
\label{tabfour}}
\begin{center}
\begin{tabular}{|c|c|}
  \hline
  i & $\Gamma_{i}$ \\
  \hline
  1 & $1/16+N/96$ \\
  2 & $1/8+N/48$ \\
  3 & 0 \\
  4 & 1/8 \\
  5 & $N/8$ \\
  6 & $1/16+1/(8N^2)$ \\
  8 & $N/16-1/(4N)$ \\
  \hline
\end{tabular}
\end{center}
\end{table}

When $N$ is arbitrary, there is in general another operator
\cite{gass}, $\textrm{tr}[L_{\mu}L_{\nu}L^{\mu}L^{\nu}]$, which
cannot be absorbed into the first three Gasser-Leutwyler operators
as it can for $N=3$ using trace relations.  For the purposes of
this paper, the additional operator and its divergent coefficient,
$L_0$, can be absorbed into $L_1$ through $L_3$ for the only
amplitudes of interest to which it contributes, $K \rightarrow \pi
\pi$ for $m_K = m_{\pi}$, and in general, for UKX.  Thus, we
absorb the $N$ dependence of $L_0$ into $L_1, L_2$ and $L_3$ in
Table 4.

\subsection{Role of the bilinear ($3,\overline{3}$) operator}

The bilinear ($3,\overline{3}$) operator is useful in removing the
power divergent coefficients to all orders in ChPT.  Recall that
the $\Delta I =1/2$ matrix elements of the four-quark operators in
general have a power divergent part. This power divergence reduces
to a quark bilinear times a momentum independent coefficient
\cite{blum}. The quark bilinear operator can be defined as in
\cite{bern},

\begin{equation}
\Theta^{(3,\overline{3})}\equiv \overline{s}(1-\gamma_5)d
\end{equation}

\noindent which is equal to
$\alpha^{(3,\overline{3})}\textrm{Tr}(\lambda_6\Sigma)$ to lowest
order in chiral perturbation theory, where in our conventions,
$\alpha^{(3,\overline{3})}=\frac{-f^2}{2}B_0$. As illustrated in
Section 6, the matrix elements of this operator can be used to
eliminate the power divergences in the effective four-quark
operator matrix elements \cite{sharptwo}.  This subtraction is to
all orders in ChPT, and in Section 6 we demonstrate this
explicitly to NLO in the partially quenched theory, following the
derivation in \cite{blum}.  It is crucial that the subtraction be
independent of ChPT, since the higher order corrections of the
power divergent operator can far exceed the physical contributions
that one is trying to calculate.  In order to carry out the
argument to NLO in (PQ)ChPT for the case of the (8,1)'s we need
the NLO LEC contribution of the $\Theta^{(3,\overline{3})}$
operator.  The effect of the subtraction involving this operator
is to eliminate the LEC, $\alpha_2$, at leading order, and to
transform the NLO (8,1) coefficients to a linear combination
involving the Gasser-Leutwyler coefficients. The chiral rotation
eliminates the power divergent scale dependence (proportional to
$\alpha_2$) of the LEC's to NLO. The effect of this transformation
on the individual coefficients is given in Table 5 \footnote{The
table is constructed using information given in \cite{kambor}}.

\begin{table}[htbp]
\caption{When the tadpole terms are subtracted via the
$\Theta^{(3,\overline{3})}$ operator, the (8,1) NLO coefficients
are transformed to new linear combinations involving the
Gasser-Leutwyler coefficients.  These new combinations no longer
have power divergences.}
\begin{center}
\begin{tabular}{|c|}
  \hline
   Transformed Coefficients \\
  \hline
    $e^r_1 \rightarrow e^r_1-(4\alpha_2/f^2) (2L^r_8+H^r_2)$ \\
    $e^r_2 \rightarrow e^r_2-(16\alpha_2/f^2) L^r_6$ \\
    $e^r_3 \rightarrow e^r_3+(4\alpha_2/f^2)(-2L^r_8+H^r_2)$ \\
    $e^r_5 \rightarrow e^r_5-(4\alpha_2/f^2) H^r_2 $ \\
    $e^r_{10} \rightarrow e^r_{10}-(4\alpha_2/f^2) L^r_5$ \\
    $e^r_{13} \rightarrow e^r_{13}-(8\alpha_2/f^2) L^r_4 $ \\
    $e^r_{15} \rightarrow e^r_{15}+(4\alpha_2/f^2) L^r_5 $ \\
  \hline
\end{tabular}
\end{center}
\end{table}

\section{$K \to \pi \pi $ without 3 momentum insertion}

In \cite{laiho} we have shown that all the amplitudes of
 interest for the (8,1)'s and (27,1)'s can be obtained
to NLO in ChPT when one uses lattice computations from $
K^{0}\rightarrow \overline{K^{0}}$, $K\to| 0 \rangle$, $K\to\pi$
with momentum and $K\to\pi\pi$ at the two unphysical kinematics
points UK1 \cite{berntwo} $\Rightarrow m_K=m_\pi$ and UK2
\cite{dawson} $\Rightarrow m_K=2m_\pi$.  Specifically, these two
points correspond to threshold, and, thereby, the Maiani-Testa
theorem is evaded \cite{maiani}.  Here we ask how far one can get
by {\it not\/} using 3-momentum insertion in $K\to\pi$ and using
only non-degenerate quarks so that on the lattice $m_K\ne m_\pi$.
  In this case one is using energy insertion
with $q^2=(m_K-m_\pi)^2$.  The motivation for this should be
clear. Not only can 3-momentum insertion add to the computational
cost, it also tends to be noisy.  On the other hand, in a typical
weak matrix element calculation, $m_K\ne m_\pi$ is relatively
inexpensive to implement, since light quarks with several masses
are needed anyway.

At $O(p^4)$ in $K\to\pi$ one can see explicitly \cite{laiho} that
different LEC's appear in front of $(p_K\cdot p_\pi)^2$ than in
front of $m^2_Km^2_\pi$.  In general $p_K\cdot p_\pi \ne
m_Km_\pi$, so it is not clear if all of the LEC's needed for
constructing $K\to\pi\pi$ to $O(p^4)$ can be obtained if one
restricts to no 3-momentum insertion in $K\to\pi$.  We find that
for all cases of interest without 3-momentum insertion, although
some low-energy constants cannot be obtained, the linear
combinations that are needed for constructing the physical NLO
amplitude can always be obtained.  This reduces the necessary
computational effort considerably. This section will be restricted
to demonstrating this result for the full theory, but in the next
section we show that the same result holds also for the partially
quenched case in the PQS framework. It is not known whether this
continues to hold in the PQN framework, which is considerably more
complicated at NLO.

In \cite{laiho} we showed how to get physical $K \to \pi \pi $
amplitudes for both $ \Delta I=1/2$ and 3/2 cases to NLO. Since $K
\to \pi$ amplitudes do not conserve four-momentum for $m_s \neq
m_d$, it is necessary to allow the weak operator to transfer a
four-momentum, $q \equiv p_K-p_\pi$, as in \cite{cirig}.  This is
also necessary for the case of $K \to \pi \pi$ at $m_K=m_\pi$
\cite{berntwo}. Our method \cite{laiho} requires computation of $K
\to \pi \pi$ at unphysical kinematics because there are low energy
constants which appear in $K \to \pi \pi$ but do not appear in $K
\to \pi$ at all \cite{bijnens,golt}.

There exist other unphysical kinematics values (besides UK1 and
UK2) for the $K\to\pi\pi$ amplitudes where the initial and final
state mesons are at rest that are accessible to lattice
calculations. These values of the kinematics bypass the
Maiani-Testa theorem because the final state pions are at
threshold, but energy insertion (or removal) at the weak operator
has to take place in order to conserve 4-momentum. These
amplitudes have no imaginary parts as long as $m_K \geq m_\pi$,
($m_s \geq m_{u,d}$, since then a two kaon intermediate state
cannot go on-shell), and so bypass the Maiani-Testa theorem. On
the lattice, so long as one studies the appropriate correlation
function as a function of Euclidean times and does not sum over
the time index, the weak operator can insert (or remove) the
necessary amount of energy. What we are calling UK1 ($m_K=m_\pi$)
and UK2 ($m_K=2 m_\pi$) are just special examples of this more
general kinematics which we call UKX, which is itself a special
case of the SPQcdR kinematics (one pion at rest, the other with
3-momentum inserted) \cite{lin} where both pion 3-momenta are
zero, and $E_\pi$, the energy of each pion, is equal to $m_\pi$.

We point out that the UKX kinematics is at threshold because of
the ability of the weak operator to inject or remove the necessary
energy, so that the Maiani-Testa theorem is bypassed even for
$\Delta I=1/2$ amplitudes.  As pointed out by \cite{linthree}, the
case where $m_K=m_\pi$ has a number of difficulties, especially in
the partially quenched theory.  It is, therefore, necessary to
consider the more general kinematics of UKX (with $m_K>m_\pi$) in
order to bypass this problem. Using UKX, one can then obtain all
of the LEC's necessary to construct the (8,1), $K\to\pi\pi$
amplitudes in both the full theory and in the partially quenched
case, if a numerical calculation at UK1 is difficult or
impossible.  We present NLO results for UKX in the partially
quenched theory in Section 8.

Finally, it is also useful to emphasize that even when one works
to LO, $K\to\pi$ with $m_K\ne m_\pi$ (without 3-momentum
insertion) suffices to give $K\to\pi\pi$ at that order, thus
providing an alternate subtraction method to the one that has been
used recently \cite{noaki,blum,bhatt,pk} with $K\to 0$
\cite{bern,maianitwo}.

\subsection{$ (27,1), \Delta I=3/2$}

The expression for the physical $K \to \pi \pi$, including only
tree level $O(p^2)$ and $O(p^4)$ weak counterterms, is
\cite{laiho}

\begin{eqnarray}\label{14}
\langle \pi^{+}\pi^{-}|{\cal O}^{(27,1),(3/2)}|K^{0}\rangle_{ct}
&=& -\frac{4i\alpha_{27}}{f_K f^{2}_\pi}(m^{2}_{K}-m^{2}_{\pi}) +
\frac{4i}{f_K f^{2}_\pi}(m^{2}_{K}-m^{2}_{\pi}) \nonumber \\ &&
\times [(-d^{r}_{4}+d_{5}^{r} -4d^{r}_{7})m^{2}_{K}
+(4d^{r}_{2}+4d^{r}_{20} \nonumber
\\ && -16d^{r}_{24}-4d^{r}_{4}-2d^{r}_{7})m^{2}_{\pi}].
\end{eqnarray}

The counterterm expressions needed to construct this physical
amplitude are given in \cite{laiho} Eqs (21--24), and the finite
logarithmic contributions are given there in Appendix C\null.
 Counterterms needed to construct the above $K \to \pi \pi$
amplitude can be obtained from $K^{0}\rightarrow
\overline{K^{0}}$; $ K^+ \rightarrow \pi^+, \Delta I =3/2$
(non-degenerate quarks); and $K \to \pi \pi, \Delta I=3/2$ at only
one value of the unphysical kinematics (e.g. UK1\footnote{For the
$\Delta I=3/2$ case there is no difficulty at UK1.}). Note that
the expression for $K^+ \to \pi^+, \Delta I=3/2$ reduces, for the
case of no 3-momentum insertion, i.e.\ $q^2=(m_K-m_\pi)^2$, to

\begin{eqnarray}\label{11}
\langle \pi^{+}|{\cal O}^{(27,1),(3/2)}|K^{+}\rangle_{ct} &=&
-\frac{4}{f^{2}} \alpha_{27}m_{K} m_{\pi}
+\frac{8}{f^{2}}[(2d^{r}_{2}-8d^{r}_{24})m^{2}_{K}m^{2}_{\pi}
\nonumber \\ && +(d^{r}_{20}
 -d^{r}_{4} -2d^{r}_{7}) m^{3}_{K}
 m_{\pi}
+(d^{r}_{20}-d^{r}_{4} \nonumber \\ && -d^{r}_{7})m_{K}
m^{3}_{\pi}],
\end{eqnarray}

The logarithmic corrections to this expression reduce to the value
given in Appendix C of this paper.  Fits to the $K \rightarrow
\pi$ data can therefore give $d^{r}_7$, $d^{r}_{20}-d^{r}_4$ and
$d^{r}_2-4d^{r}_{24}$. Using these in the $K \to \pi \pi$
amplitude at the unphysical kinematics point (UK1) $m_K=m_\pi=m$
(Eq (23) of \cite{laiho}) gives $d^r_4-d^r_5$.  The four linear
combinations
$[d^{r}_2-4d^r_{24},d^r_7,d^r_4-d^r_5,d^r_4-d^r_{20}]$ are
sufficient to determine $K \to \pi \pi$, $\Delta I=3/2$ at the
physical kinematics as given in Eq.(34).  Comparing Eq.(35) with
the more general case of 3-momentum insertion, Eq.(22) of
\cite{laiho}, we see that the latter allows for separate
determinations of $d^r_2$ and $d^r_{24}$, whereas the simpler case
of $m_K\ne m_\pi$ without 3-momentum insertion, Eq.(35), gives
only the linear combination $d^r_2 - 4d^r_{24}$.  Nevertheless,
that suffices to get the job done.

\subsection{$ (8,1)+(27,1), \Delta I=1/2$}

Recall that this is the most complicated case. The counterterms
necessary to construct $O(p^4)$, $[(8,1)+(27,1)]$, $ \Delta
I=1/2$, $K \to \pi \pi$ amplitudes relevant for operators such as
$Q^{1/2}_2$, which are mixed, can be obtained from the above
values for $d^r_{i}$'s and from the following $\Delta I=1/2$
processes: $ K^{0} \rightarrow 0; K^{+}\rightarrow \pi^{+}, \Delta
I=1/2 $ (non-degenerate quarks); and $ K \rightarrow \pi \pi,
\Delta I=1/2 $ at two unphysical kinematics. All of the needed
counterterm amplitudes appear in Section 4b of \cite{laiho}, and
the corresponding logarithmic corrections appear in Appendix D of
that paper.  Note that an error was discovered since publication
of that work in Eq (31) and in Appendix D, Eq (D6).  The correct
expressions appear here in Appendix F. Again, it is sufficient to
allow $q^2=(m_K-m_\pi)^2$ in the expression for $K \to \pi$,
\cite{laiho} Eqs (28) and (29).  These equations become

\begin{eqnarray}\label{17}
\langle \pi^{+}|{\cal O}^{(27,1),(1/2)}|K^{+}\rangle_{ct}& =&
-\frac{4}{f^{2}}\alpha_{27}m_{K} m_{\pi}  -
\frac{8}{f^{2}}[6d^{r}_{1}m^{4}_{K} \nonumber \\ &&
+(-6d^{r}_{1}-2d^{r}_{2}+8d^{r}_{24})m^{2}_{K} m^{2}_{\pi} \nonumber \\
&& + (-d^{r}_{20}+d^{r}_{4}-3d^{r}_{6}+2d^{r}_{7})m^{3}_{K}
m_{\pi} \nonumber \\ && +
(-d^{r}_{20}+d^{r}_{4}+3d^{r}_{6}+d^{r}_{7})m_{K} m^{3}_{\pi}],
\end{eqnarray}

\begin{eqnarray}\label{18}
\langle \pi^{+}|{\cal O}^{(8,1)}|K^{+}\rangle_{ct} &=&
\frac{4}{f^{2}}\alpha_{1}m_{K} m_{\pi}-
\frac{4}{f^{2}}\alpha_{2}m^{2}_{K} -
\frac{8}{f^{2}}[2(e^{r}_{1}+e^{r}_{2}-e^{r}_{5})m^{4}_{K}\nonumber
\\ && + (e^{r}_{2}+2e^{r}_{3} + 2e^{r}_{5}-8e^{r}_{39})
m^{2}_{K}m^{2}_{\pi}
 + (2e^{r}_{35}-2e^{r}_{10})m^{3}_{K} m_{\pi} \nonumber \\ && +
(2e^{r}_{35}-e^{r}_{11})m_{K} m^{3}_{\pi}],
\end{eqnarray}

The logarithmic corrections associated with the above two
amplitudes are given in Appendix C of this paper.  In evaluating,
for example, $\bra\pi^+|Q^{1/2}_2 |K^+\ket$, the right hand sides
of Eqs. (36) and (37) have to be added. In fitting to lattice
data, for example, the $m^4_K$ coefficient would give the
combination $[6d^r_1 +2e^r_1 +2e^r_2 -2e^r_5]$. Also, in comparing
Eq. (37) with Eq. (29) in \cite{laiho} without 3-momentum
insertion, one can no longer separately obtain
$(e^r_2+2e^r_3+2e^r_5)$ and $-8e^r_{39}$ but only their sum;
however, this is again sufficient to obtain the physical $ K \to
\pi \pi $ amplitudes [\cite{laiho} Eqs (34),(35)] to NLO\null.

We point out that for the $\Delta I =1/2$ amplitudes there are
power divergences that must be subtracted using the
$\Theta^{(3,\overline{3})}$ operator introduced at the end of
Section 4.  It is crucial that the subtraction is to all orders in
ChPT, since the higher order corrections of the power divergent
operator can far exceed the physical contributions that one is
trying to determine. This is discussed in more detail in Sections
7 and 8 for the partially quenched case, where we follow the
derivation in \cite{blum}, given for the leading order case in the
full theory (although there the analysis was done with quenched
data). The result of the subtraction is to eliminate the power
divergent coefficient, $\alpha_2$, and to transform the (8,1) NLO
LEC's to the values given in Table 5. Thus, fits to the subtracted
lattice data will give the transformed coefficients, where their
power divergences have been eliminated. This is what we want,
since only these finite combinations appear in physical
quantities.  The process described in the above discussion on the
determination of the NLO LEC's, along with that in \cite{laiho},
is not invalidated.

One can determine, using the $ \Delta I=3/2$ amplitudes, the
following constants:
$[d^r_1,d^{r}_2-4d^r_{24},d^r_7,d^r_4-d^r_5,d^r_4-d^r_{20}]$. Here
$d^r_1$ and $d^r_7$ can both be determined from $K \to \bar{K}$
[\cite{laiho}, Eq(21)], and the procedure for the others is given
in the previous section.  Given these, one can obtain
$e^r_{2,rot}$ and $e^r_{1,rot}-e^r_{5,rot}$ from $K^0 \to 0$. Note
that the values of the coefficients obtained are those of the
subtracted amplitudes, and that the subscript refers to the LEC
after the chiral rotation of Table 5 has been performed. Only
after the subtraction can one fit to the lattice data using ChPT.
Given the previous information one can obtain
$e^r_{1,rot}+e^r_{3,rot}-4e^r_{39}$,
$e^r_{10,rot}-e^r_{35}+\frac32d^r_6$, and
$2e^r_{10,rot}-e^r_{11}+6d^r_6$ from Eqs (36) and (37), after the
subtraction has been performed. From Eqs (30) and
(31)\footnote{Note that Eq (31) of \cite{laiho} is corrected in
Appendix F, but this does not change the conclusion here.} of
reference \cite{laiho} for $K \to \pi \pi$, $m_K=m_\pi=m$
(UK1\footnote{Although \cite{linthree} have pointed out that UK1
may be computationally demanding even for the full theory, it is
not ruled out.  In any case, for extracting the LEC's one can use
the more general kinematics which we call UKX, as discussed
earlier in this section.}), one can then obtain
$e^r_{11}+2e^r_{15,rot}-3d^r_6$. Making use of all of the input
thus obtained into Eqs (32) and (33) of reference \cite{laiho} for
$K \to \pi \pi$, $m_K=2m_\pi$ (UK2), yields
$e^r_{13,rot}-\frac32d^r_6$ (after the subtraction). Thus, the 11
linear combinations necessary to construct the physical $K \to \pi
\pi$ at NLO (without using 3-momentum insertion but with
non-degenerate quarks in $K\to\pi$) are $[d^r_1, d^r_2-4d^r_{24},
d^r_7, d^r_4-d^r_5, d^r_4-d^r_{20}, e^r_{2,rot},
e^r_{1,rot}+e^r_{3,rot}-4e^r_{39}, e^r_{10,rot}-e^r_{35} +\frac32
d^r_6, 2e^r_{10,rot}-e^r_{11}+6d^r_6, e^r_{11}+2e^r_{15,rot} -3
d^r_6, e^r_{13,rot}-\frac32d^r_6]$.

\subsection{(8,1)}

The case of pure (8,1) operators, e.g., $Q_6$, is simpler than the
previous case of mixed $\Delta I =1/2$ operators, and is
phenomenologically the most important one as it gives the dominant
contribution to the CP-odd phase of $\epe$ coming from
QCD-penguins. For this case the six needed linear combinations are
$[e^r_{2,rot}, e^r_{1,rot}+e^r_{3,rot}-4e^r_{39},
e^r_{35}-e^r_{10,rot}, 2e^r_{35}-e^r_{11}, e^r_{11}+2e^r_{15,rot},
e^r_{13,rot}]$.  The first of these  is obtained from $K\to 0$.
The second requires both $K\to 0$ and $K\to\pi$ ($m_K \neq
m_\pi$). The third and fourth are also obtained from $K\to\pi$.
$K\to\pi\pi$ at UK1 then gives the fifth, and $K\to \pi\pi$ at UK2
gives the sixth coefficient.  Since it is likely that UK1 will
prove to be particularly difficult \cite{linthree}, it is possible
to use another set of allowed values of UKX in order to obtain the
remaining coefficients. Of course, one will want to do such a
calculation using UKX anyway for the additional redundancy.  All
LEC's are those that would be obtained from a fit to lattice data
after the power divergent subtraction has been performed.

\subsection{(8,8)}

Since the leading order (8,8) begins at $O(p^0)$, the NLO
contribution comes at $O(p^2)$.  As an example, Eq (36) from
\cite{cirig} is given (with our normalization of $f$ and our
convention for the $c_i$'s),

\begin{eqnarray}\label{11b}
\langle \pi^{0}|{\cal O}^{(8,8)}|K^{0}\rangle_{ct}=
\frac{2\sqrt{2}}{f^{2}}\left[-\left(\frac13c_1+c_2+\frac23c_3\right)p_K
\cdot p_\pi
 -\frac23c_4m^2_K\right].
\end{eqnarray}

\noindent Now with $m_K\ne m_\pi$, even when both mesons are at
rest, and $p_K\cdot p_\pi= m_Km_\pi$, there is no loss of
information, and all the coefficients can be obtained at NLO
without 3-momentum insertion.

\section{Calculating $K \to \pi \pi $ Amplitudes in PQChPT}

In this section we discuss the ambiguity of PQChPT in the $\Delta
I=1/2$ case where eye-diagrams appear.  At least two ways arise in
the context of PQChPT for dealing with the gluonic penguins, the
PQS and the PQN methods.  These are described, and their
predictions at leading order are compared using formulas given by
Golterman and Pallante.  In the following subsections we give NLO
expressions in PQChPT for the ingredients necessary to obtain $K
\rightarrow \pi \pi$ at $O(p^2)$ and $O(p^4)$ for the (8,8)'s and
(8,1)'s, respectively. For the (8,8)'s it is necessary to know $K
\rightarrow \pi$, $\Delta I=3/2$ and 1/2 in order to get all the
coefficients at NLO, as shown in \cite{cirig}.  This remains true
in PQChPT. The important point to note is that one can construct
$K\to\pi\pi$ amplitudes for the (8,8) operator to NLO using only
$K\to\pi$ with degenerate quark masses ($m_K=m_\pi$), along with
$K\to0$ to perform the $\Delta I=1/2$ power subtraction.

For the (8,1)'s, one needs $K \rightarrow 0$, $K \rightarrow \pi$
with non-degenerate quarks, and $K \rightarrow \pi \pi$ at two
values of unphysical kinematics, e.g. $m_K = m_{\pi}$ (UK1) and
$m_K=2m_{\pi}$ (UK2), as shown in \cite{laiho} in full ChPT to
NLO.  We have also introduced in Section 5 the kinematics for
$K\to\pi\pi$ accessible to the lattice which we have called UKX,
of which UK1 and UK2 are special cases.  Reference \cite{linthree}
has demonstrated that UK1 has difficulties in the full theory, and
is not tractable in the partially quenched theory due to enhanced
finite volume effects. One can still obtain all of the needed
LEC's to construct $K\to\pi\pi$ to NLO for the (8,1)'s from UKX,
however. This remains true in PQChPT only when one is working
within the PQS framework. This paper, therefore, follows the
prescription of the PQS method for the (8,1)'s.  Note that in the
PQN method it is not clear if all the ingredients needed for
constructing the physical $K\to\pi\pi$ amplitudes to NLO can be
determined from the lattice, except in the full theory ($N=3$,
$m_{sea}=m_{val}$) where the two methods coincide.  Note, also,
that the (8,1), $K \rightarrow \pi \pi$ amplitudes at UKX are
afflicted by enhanced finite volume corrections except when
$m_{sea}=m_u=m_d$ for both the PQS and PQN methods.

In the PQChPT case (as in full ChPT) the $K \rightarrow \pi$
amplitudes require non-degenerate quarks, $m_s \neq m_u=m_d$, in
order to extract all of the necessary LEC's from them.  Since this
amplitude does not conserve four-momentum, for $m_s \neq m_d$ the
weak operator must transfer a four momentum $q\equiv p_K-p_\pi$.
The conclusion of the previous section that three-momentum
insertion is not essential holds also in the case of PQChPT.

The diagrams to be evaluated for the NLO corrections are shown in
Fig 1.  The topologies are unchanged from \cite{laiho}, although
additional pseudo-fermion ghost and sea meson fields propagate in
the loops.  The renormalization of the external legs via the
strong interaction must be taken into account.

\begin{figure}[htbp]
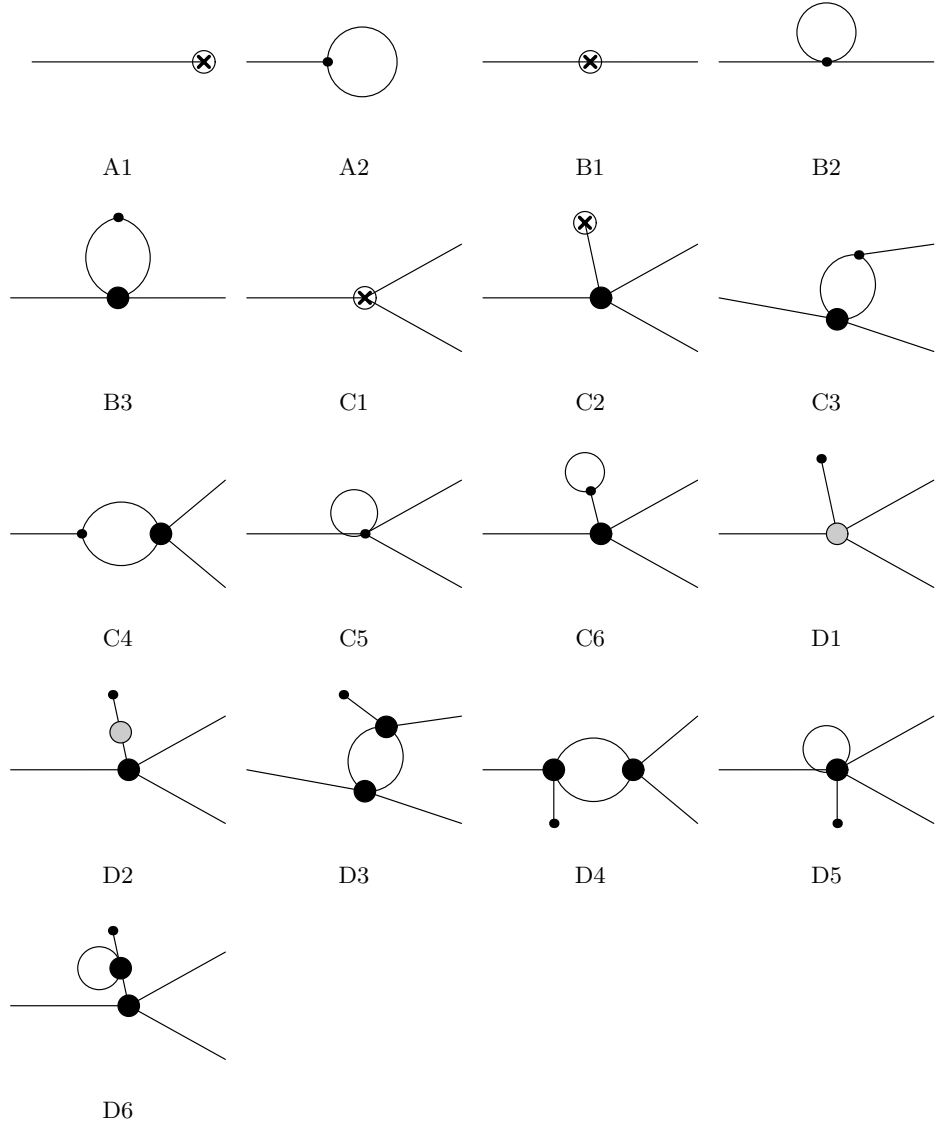


\unitlength=1bp%

\begin{feynartspicture}(432,445)(4,5)

\FADiagram{A1} \FAProp(2.,10.)(18.,10.)(0.,){/Straight}{0}
\FAVert(18.,10.){2}

\FADiagram{A2} \FAProp(0.,10.)(7.5,10.)(0.,){/Straight}{0}
\FAProp(7.5,10.)(7.5,10.)(14.,10.){/Straight}{0}
\FAVert(7.5,10.){0}

\FADiagram{B1} \FAProp(0.,10.)(10.,10.)(0.,){/Straight}{0}
\FAProp(20.,10.)(10.,10.)(0.,){/Straight}{0} \FAVert(10.,10.){2}

\FADiagram{B2} \FAProp(0.,10.)(10.,10.)(0.,){/Straight}{0}
\FAProp(20.,10.)(10.,10.)(0.,){/Straight}{0}
\FAProp(10.,10.)(10.,10.)(10.,15.5){/Straight}{0}
\FAVert(10.,10.){0}

\FADiagram{B3} \FAProp(0.,10.)(10.,10.)(0.,){/Straight}{0}
\FAProp(20.,10.)(10.,10.)(0.,){/Straight}{0}
\FAProp(10.,17.5)(10.,10.)(-0.8,){/Straight}{0}
\FAProp(10.,17.5)(10.,10.)(0.8,){/Straight}{0}
\FAVert(10.,10.){-5} \FAVert(10.,17.5){0}

\FADiagram{C1} \FAProp(0.,10.)(11.,10.)(0.,){/Straight}{0}
\FAProp(20.,15.)(11.,10.)(0.,){/Straight}{0}
\FAProp(20.,5.)(11.,10.)(0.,){/Straight}{0} \FAVert(11.,10.){2}

\FADiagram{C2} \FAProp(0.,10.)(11.,10.)(0.,){/Straight}{0}
\FAProp(20.,15.)(11.,10.)(0.,){/Straight}{0}
\FAProp(20.,5.)(11.,10.)(0.,){/Straight}{0}
\FAProp(9.5,17.)(11.,10.)(0.,){/Straight}{0} \FAVert(11.,10.){-5}
\FAVert(9.5,17.){2}

\FADiagram{C3} \FAProp(0.,10.)(11.,8.)(0.,){/Straight}{0}
\FAProp(20.,15.)(13.,14.)(0.,){/Straight}{0}
\FAProp(20.,5.)(11.,8.)(0.,){/Straight}{0}
\FAProp(13.,14.)(11.,8.)(0.8,){/Straight}{0}
\FAProp(13.,14.)(11.,8.)(-0.8,){/Straight}{0} \FAVert(13.,14.){0}
\FAVert(11.,8.){-5}

\FADiagram{C4} \FAProp(0.,10.)(6.6,10.)(0.,){/Straight}{0}
\FAProp(20.,15.)(14.,10.)(0.,){/Straight}{0}
\FAProp(20.,5.)(14.,10.)(0.,){/Straight}{0}
\FAProp(6.6,10.)(14.,10.)(0.8,){/Straight}{0}
\FAProp(6.6,10.)(14.,10.)(-0.8,){/Straight}{0} \FAVert(6.6,10.){0}
\FAVert(14.,10.){-5}

\FADiagram{C5} \FAProp(0,10.)(11.,10.)(0,){/Straight}{0}
\FAProp(20.,15.)(11.,10.)(0,){/Straight}{0}
\FAProp(20.,5.)(11.,10.)(0,){/Straight}{0}
\FAProp(11.,10.)(11.,10.)(9.0034,13.8721){/Straight}{0}
\FAVert(11.,10.){0}

\FADiagram{C6} \FAProp(0.,10.)(11.,10.)(0.,){/Straight}{0}
\FAProp(20.,15.)(11.,10.)(0.,){/Straight}{0}
\FAProp(20.,5.)(11.,10.)(0.,){/Straight}{0}
\FAProp(10.,14.)(11.,10.)(0.,){/Straight}{0}
\FAProp(10.,14.)(10.,14.)(9.,17.5){/Straight}{0}
\FAVert(11.,10.){-5} \FAVert(10.,14.){0}

\FADiagram{D1} \FAProp(0.,10.)(11.,10.)(0.,){/Straight}{0}
\FAProp(20.,15.)(11.,10.)(0.,){/Straight}{0}
\FAProp(20.,5.)(11.,10.)(0.,){/Straight}{0}
\FAProp(9.5,17.)(11.,10.)(0.,){/Straight}{0} \FAVert(11.,10.){-1}
\FAVert(9.5,17.){0}

\FADiagram{D2} \FAProp(0,10.)(11.,10.)(0,){/Straight}{0}
\FAProp(20.,15.)(11.,10.)(0,){/Straight}{0}
\FAProp(20.,5.)(11.,10.)(0,){/Straight}{0}
\FAProp(9.5,17)(10.25,13.5)(0,){/Straight}{0}
\FAProp(10.25,13.5)(11.,10.)(0,){/Straight}{0}
\FAVert(11.,10.){-5} \FAVert(9.5,17.){0} \FAVert(10.25,13.5){-1}

\FADiagram{D3} \FAProp(0.,10.)(11.,8.)(0.,){/Straight}{0}
\FAProp(20.,15.)(13.,14.)(0.,){/Straight}{0}
\FAProp(20.,5.)(11.,8.)(0.,){/Straight}{0}
\FAProp(13.,14.)(11.,8.)(0.8,){/Straight}{0}
\FAProp(13.,14.)(11.,8.)(-0.8,){/Straight}{0}
\FAProp(13.,14.)(9.,17.)(0.,){/Straight}{0} \FAVert(13.,14.){-5}
\FAVert(11.,8.){-5} \FAVert(9.,17.){0}

\FADiagram{D4} \FAProp(0.,10.)(6.6,10.)(0.,){/Straight}{0}
\FAProp(20.,15.)(14.,10.)(0.,){/Straight}{0}
\FAProp(20.,5.)(14.,10.)(0.,){/Straight}{0}
\FAProp(6.6,10.)(14.,10.)(0.8,){/Straight}{0}
\FAProp(6.6,10.)(14.,10.)(-0.8,){/Straight}{0}
\FAProp(6.6,10.)(6.6,5.)(0.,){/Straight}{0} \FAVert(6.6,10.){-5}
\FAVert(14.,10.){-5} \FAVert(6.6,5.){0}

\FADiagram{D5} \FAProp(0,10.)(11.,10.)(0,){/Straight}{0}
\FAProp(20.,15.)(11.,10.)(0,){/Straight}{0}
\FAProp(20.,5.)(11.,10.)(0,){/Straight}{0}
\FAProp(11.,10.)(11.,10.)(9.0034,13.8721){/Straight}{0}
\FAProp(11.,10.)(11.,5.)(0.,){/Straight}{0} \FAVert(11.,10.){-5}
\FAVert(11.,5.){0}

\FADiagram{D6} \FAProp(0,10.)(11.,10.)(0,){/Straight}{0}
\FAProp(20.,15.)(11.,10.)(0,){/Straight}{0}
\FAProp(20.,5.)(11.,10.)(0,){/Straight}{0}
\FAProp(9.5,17)(10.25,13.5)(0,){/Straight}{0}
\FAProp(10.25,13.5)(11.,10.)(0,){/Straight}{0}
\FAProp(10.25,13.5)(10.25,13.5)(6.25,13.5){/Straight}{0}
\FAVert(11.,10.){-5} \FAVert(9.5,17.){0} \FAVert(10.25,13.5){-5}

\end{feynartspicture}

\caption{Diagrams needed to evaluate the NLO amplitudes in
(PQ)ChPT.  NLO corrections include tree-level diagrams with
insertion of the NLO weak vertices (crossed circles), tree-level
diagrams with insertion of $O(p^{4})$ strong vertices (lightly
shaded circles), one-loop diagrams with insertions of the LO weak
vertices (small filled circles) and the $O(p^{2})$ strong vertices
(big filled circles). The lines represent the propagators of
mesons comprised of valence, ghost, and sea quarks. A1 and A2 are
for $K\to 0$. B1-B3 are for $K\to\pi$. C1-C6 and D1-D6 are for
$K\to\pi \pi$.}
\end{figure}

\subsection{The Treatment of Eye Graphs}

There is a subtlety concerning the $\Delta I=1/2$ amplitudes in
the partially quenched theory, and this has been discussed by
Golterman and Pallante for the case of the gluonic penguins
\cite{goltthree, golt, goltfour}.  What follows is a summary of
their work.  To illustrate the subtlety, we discuss the situation
for the $Q_6$ gluonic penguin operator, given by

     \begin{equation}
Q_{6}=\overline{s}_{a} \gamma_{\mu} (1-\gamma^{5}) d_{b} \sum_{q}
    \overline{q}_{b}\gamma^{\mu} (1+\gamma^{5}) q_{a}.
    \end{equation}

\noindent The right part of this operator is a sum over light
flavors, $q=u, d, s$, so in the full theory the right hand part is
a flavor singlet under the symmetry group $\textrm{SU}(3)_R$. In
the partially quenched theory one has at least two options. One
may choose to sum over all the quarks, including sea and ghost in
which case the right component of the operator transforms as a
singlet under the extended symmetry group; therefore, this is
called the PQS (partially quenched singlet) option.  In the second
option, one may choose to sum in Eq (39) over only the valence
quarks.  In this case the operator is a linear combination of two
terms, one of which transforms as a singlet under the extended
symmetry group, while the other does not transform as a singlet
under the irreducible representation of the extended symmetry
group (rather, for $Q_6$, it transforms in the adjoint
representation); therefore, we choose to call this the PQN
(partially quenched non-singlet) method.

Given that the flavor blind, vector character of the
quark-quark-gluon elementary interaction in QCD plays a crucial
role in leading to the explicit singlet form [Eq (39)] of the
right-hand part of the penguin operator, it seems reasonable to
preserve this basic character in generalization to the partially
quenched case which contains additional quarks. This provides the
rationale for the PQS option.

The origin of the PQN option is quite different; it is, in fact,
the straightforward implementation of the quenched approximation
to a lattice calculation of the necessary Green's functions.  The
usual practice leads one to use only the valence quarks in the
necessary Wick contractions for, say $\langle \pi |Q_6|K\rangle$,
which then lead to valence quark loops (see Fig 2a,b), the
so-called eye graphs.  In such an implementation all other quark
loops are computed when the fermion determinant is evaluated in
the generation of the gauge configurations.  When one partially
quenches in the PQN method, the gauge configurations are generated
using the number and mass of the sea quarks, but the propagators
for the loops of the eye graphs (Fig 2a,b) are still computed with
those of the valence quarks.  In the partially quenched case where
the sum in Eq (39) is over the valence quarks only, as mentioned
above, the operator is a linear combination of two terms, only one
of which transforms as a singlet under the extended symmetry
group.

Fig 2 shows the Green's function relevant for a lattice evaluation
of $\langle \pi |Q_6|K\rangle$ consisting of the two eye graphs
originating from the Wick contractions.  Any number of gluon lines
from the background gauge configurations (not explicitly shown)
are understood in such a pictorial representation of these
non-perturbative graphs.  As usual, one of the Wick contractions
is a product of two color traces (Fig 2a), while the second is a
single trace over color indices.  In the PQS implementation of the
$Q_6$ penguin operator, in the quenched case where $q\overline{q}$
loops in the gluon propagation are not included, the eye graph
(Fig 2b) with a single color trace should also be excluded, for
consistency \cite{goltthree,goltfour}.

In the PQN option of calculating $\langle \pi |Q_6|K\rangle$, one
uses valence quarks for the propagators of the eye-graphs in the
corresponding Green's function, as this appears analogous to the
usual practice in lattice computations.  However, the situation at
hand demands caution.  Lattice calculation of $\langle \pi
|Q_6|K\rangle$ is qualitatively different in important aspects
from (say) spectrum, decay-constant or form-factor calculations.

To trace the potential inconsistency we show the weak operator
with a magnified view in the non-perturbative eye-graph (Fig 3).
Inside the dashed lines is the magnified short distance effective
penguin operator; outside of these dashed lines any number of soft
gluon lines from the background gauge configurations are
understood, just as in Fig 2. For $\langle \pi |Q_6|K\rangle$, Fig
3a and 3b correspond to the product of two color traces (Fig 2a)
and Fig 3c corresponds to the single trace over color indices (Fig
2b).  Fig 3c shows clearly that the corresponding Wick contraction
(single trace over color indices for $Q_6$, i.e., Fig 2b) in a
lattice evaluation of $\langle \pi |Q_6|K\rangle$ contains a
$q\overline{q}$ loop in the propagation of the gluon, and since in
the quenched case these are being dropped from the background
gauge configurations one may wish to exclude Fig 2b (for $Q_6$) in
the quenched approximation.  In a similar vein, for the partially
quenched case one may, for consistency, take the quark loop in the
eye graph of Fig 2b (Fig 3c) to be that of sea quarks only
\cite{goltthree,goltfour}, as in the PQS method.  This lack of
consistent (partial) quenching causes the low energy dynamics of
(P)QChPT to change between PQS and PQN methods.  It is not clear
if the additional (partially) quenched non-singlet terms that
modify the low energy dynamics correctly account for the otherwise
neglected loop contractions, or if the (partially) quenched low
energy constants from the singlet operator alone substituted into
the full ChPT formulas for $K\to\pi\pi$ provide a better estimate
for the physical amplitudes.  Thus, the appearance of eye-diagrams
has created an ambiguity because the contraction of Fig 3c yields
a quark vacuum bubble, and it is not obvious whether the
propagators to be contracted should be the sea or the valence;
again, the first choice corresponds to PQS and the second to PQN.

The correspondence between the traditional form of the
non-perturbative eye graphs as shown in Fig 2 and the
non-perturbative eye graphs with the magnified view of the penguin
operator as shown in Fig 3 for all penguin operators is as
follows. For the $Q_3$ and $Q_5$ operators, the color contraction
of Fig 2a corresponds to Fig 3c, while the color contraction of
Fig 2b corresponds to Figs 3a and 3b.  For the $Q_4$ and $Q_6$
operators, the color contraction of Fig 2a corresponds to Figs 3a
and 3b, while the color contraction of Fig 2b corresponds to Fig
3c.  For the electroweak penguins, the picture in Fig 3 carries
over, but with the gluons replaced by a photon or a Z.  In that
case, for $Q_7$ the color contraction of Fig 2a corresponds to Fig
3c, while the color contraction of Fig 2b corresponds to Figs 3a
and 3b. For $Q_8$, the color contraction of Fig 2a corresponds to
Figs 3a and 3b, while the color contraction of Fig 2b corresponds
to Fig 3c.  In short, Fig 2a corresponds to Fig 3c for the
operator $Q_i$, $i=3-8$, $i$ odd, while Fig 2b corresponds to Fig
3c for $i$ even.

\begin{figure}
\begin{center}
\includegraphics[scale=.7]{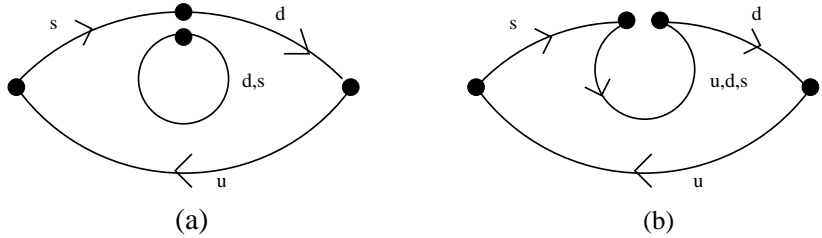}
\end{center}

\caption{The quark contractions needed for $K \rightarrow \pi$,
$\Delta I=1/2$ matrix elements include the above eye diagrams. A
connected line represents a trace over color indices, so Fig 2a
represents a product of two color traces, whereas Fig 2b
represents a single color trace.}

\end{figure}

The treatment of ChPT for the case when only valence quarks are
contracted in the eye-diagrams was first discussed by
\cite{goltthree}, for the case of the gluonic penguins.  When one
includes only the valence propagators in the eye-diagrams (no
partial quenching of the effective operator) for the case of the
gluonic penguins the right hand part of the (8,1)'s is no longer a
singlet, and there is a contribution from a non-singlet operator.
For the left-left gluonic penguins, $Q_3$ and $Q_4$, these
non-singlet contributions do not occur until next-to-leading order
\cite{goltfour}.  For the left-right gluonic penguins, $Q_5$ and
$Q_6$, the non-singlet operator transforms under the same
irreducible representation as the (8,8) electroweak penguins.
Since the (8,8)'s are NLO at $O(p^2)$, even the leading order
gluonic penguins can have logarithmic contributions from the one
loop insertions of the lowest order (8,8) operator.  These were
calculated in \cite{goltfour} for the left-right gluonic penguins,
$Q_5$ and $Q_6$.  Since the amplitudes in this case no longer
transform as pure (8,1)'s, but pick up a contribution from the
(8,8)'s, this calculation corresponds to the PQN method. It is
useful to compare the (PQ)ChPT expressions for the PQS and PQN
methods, and the next section compares the two methods at leading
order for the left-right gluonic penguins, using expressions
derived by Golterman and Pallante \cite{goltthree,golt,goltfour}.

\begin{figure}
\begin{center}
\includegraphics[scale=.7]{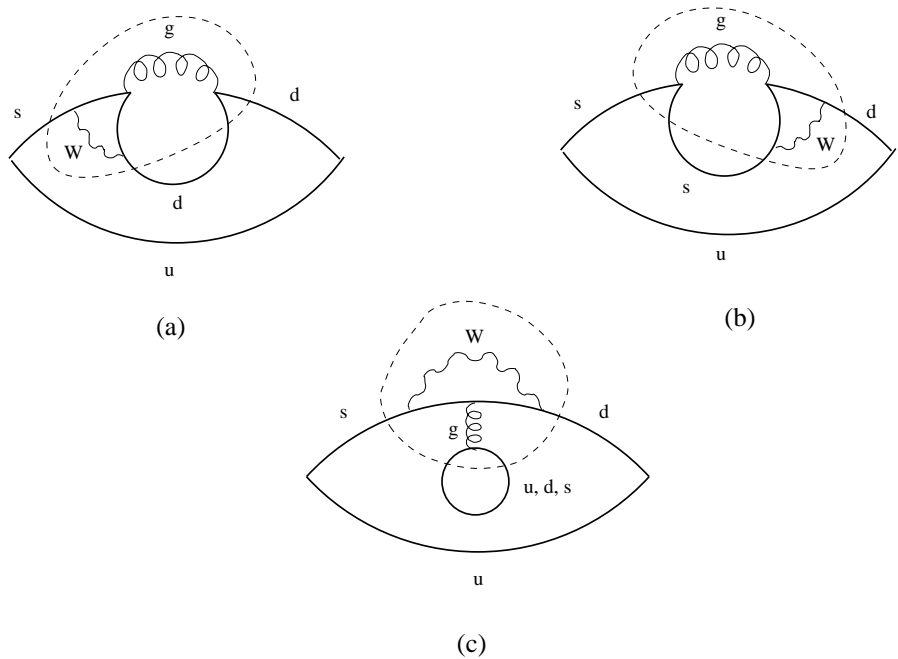}
\end{center}
\caption{The quark contractions needed for $K \rightarrow \pi$,
$\Delta I=1/2$ matrix elements include the above eye diagrams. The
weak operator is shown with a magnified view inside the dashed
lines so that one can see how it arises in perturbation theory for
the gluonic penguins. Fig 3a corresponds to $\overline{q}$ being
contracted with $d$, while Fig 3b corresponds to $q$ being
contracted with $\overline{s}$. Fig 3c corresponds to $q$ being
contracted with $\overline{q}$. For the electroweak penguins, one
would replace the gluon lines with those of photons or Z's.}

\end{figure}

We choose to work within the framework of \cite{golt}, where the
PQS method was (implicitly) used.  In this case there is the
possibility of determining the LEC's to NLO.  For the left-right
gluonic penguins, for example, at NLO in the PQN method there are
many more LEC's that appear in the amplitudes we are considering
than in the PQS method.  These are the $O(p^4)$ LEC's of the (8,8)
NNLO local operators, and it is not even clear whether one can
determine the correct linear combinations of the new LEC's
necessary to construct $K \rightarrow \pi \pi$ at NLO in ChPT from
the PQN method, except when $N=3$ and $m_{sea}=m_{val}$ (i.e.,
full QCD), as in that special situation the two options coincide.
On the other hand, for the PQS method, no new ingredients are
needed over the ones listed in our previous work \cite{laiho}
which were needed for the case of full ChPT \footnote{Note that
\cite{linthree} have shown that there are difficulties at what we
call UK1 ($K\to\pi\pi$ with $m_K=m_\pi$). There is, however, an
additional set of kinematics points that bypass the Maiani-Testa
theorem, creating the two pion state at threshold with energy
carried by the weak operator which we call UKX.  If UK1 proves
difficult or impossible one must supplement the other ingredients
with a calculation of $K\to\pi\pi$ at UKX with $m_K>m_\pi$. }. The
PQS method can also be applied to obtain all of the needed LEC's
to construct $K\to\pi\pi$ to NLO for the case $N=2$ (using the
same ingredients as for the full theory), though in this case the
LEC's are not necessarily the same as in the $N=3$ physical case.

To reiterate, in general, the PQN method is complicated by the
contributions of many more LEC's, and it is not known whether this
method can be used to NLO. Such a determination would require a
two-loop calculation. The PQS method gives us everything we need,
and is the \emph{only} method where we have demonstrated that it
is possible to obtain $K \to\pi\pi$ to NLO in ChPT. Thus, we use
the PQS prescription.

As discussed in \cite{goltthree}, the NLO (8,8) LEC's that appear
in linear combinations with the LO (8,1) LEC's in the PQN
expressions for the amplitudes of the left-right gluonic penguins
are not present for the PQS method.  For the case where $N=3$,
there is no ambiguity, and one must extract the (8,1) LEC's
separately since these LEC's take the same values as in the full
theory. However, when $N$ is \emph{not} equal to 3, it may be that
the additional (8,8) LEC's appearing in linear combinations with
the (8,1) LEC's bring the $N\neq3$ values of the (8,1) LEC's to
closer agreement with the $N=3$ values of the real world. As long
as an explicit $N=3$ lattice calculation is lacking, it may be
useful to compare the determinations of both PQN and PQS leading
order LEC's at other values of $N$ in order to learn something of
the size of the systematic error due to partial quenching
\cite{goltthree,goltfour}.  This is discussed further for
$Q_{5,6}$, LO $K\to\pi\pi$ amplitudes in the next subsection.

This paper requires $K \rightarrow 0$, $K \rightarrow \pi$, $m_s
\neq m_d=m_u$, and $K \rightarrow \pi \pi$ at two unphysical
kinematics in order to construct the physical (8,1), $K
\rightarrow \pi \pi$ amplitude for any gluonic penguin operator
($Q_{3,4,5,6}$) using the PQS prescription. Reference \cite{golt}
presented $K \rightarrow 0$, and $K \rightarrow \pi$,
$m_s=m_u=m_d$, and we agree with those calculations in the case we
consider, namely the partially quenched case with $m_{val},m_{sea}
\ll m_{\eta'}$. We extend these calculations to include all
amplitudes needed to obtain the (8,1) LEC's necessary to construct
$K \rightarrow \pi \pi$ to NLO.

In the case of the (8,8), $\Delta I=1/2$ amplitudes, one must also
make this choice of whether to (partially) quench the right side
of the penguin operator. In this case, however, the difference
comes in the choice of the quark charge matrix, $Q$. If we choose
the ghost quark charges to be equal to the valence quark charges,
then we quench the electroweak penguins, and one should ignore the
valence contributions to Fig 3c (with the gluon replaced by a
photon or Z, Fig 3c corresponds to Fig 2a for $Q_7$ and to Fig 2b
for $Q_8$) in the lattice calculation. However, if one chooses the
ghost quarks to have zero charge then the electroweak interaction
remains unquenched, and one must include the valence quarks in Fig
3c in the lattice calculation. In both cases the sea quark loop
contributions to Fig 3c vanish if we assume degenerate sea quark
masses and that the sum of the sea quark charges is zero. The
logarithmic expressions resulting from either choice for the
(8,8)'s are presented in Appendix D.

To summarize, our calculation for the (8,1) gluonic penguin matrix
elements corresponds to the PQS method.  In the corresponding
lattice calculation, the eye contractions of Fig 3c (corresponding
to Fig 2a for $Q_3$, $Q_5$, and $Q_7$, and to Fig 2b for $Q_4$,
$Q_6$ and $Q_8$) include only the sea quarks. That is, the
propagator of the internal loop of Fig 3c is calculated with the
masses of the sea quarks, not the valence quarks.  As discussed
above, when $N=3$, this, the PQS method, allows for the only known
implementation of the reduction method for the gluonic penguins.
It greatly simplifies the LO analysis \cite{goltfour}, and makes
possible a NLO determination of all of the necessary LEC's, as
demonstrated in this paper. For the (8,8) electroweak penguin
matrix elements, for degenerate sea quark masses the eye graph of
Fig 3c (with the gluon replaced by a photon) vanishes for any
number of dynamical flavors by construction (see Sect 4). Whether
one chooses to include valence quarks in the loop of Fig 3c does
not significantly alter the situation in PQChPT, and formulas for
both implementations are given in this paper.

\subsection{PQS vs PQN at Leading Order}

This section is a review of Golterman and Pallante's
\cite{goltthree,golt,goltfour} results for the leading order,
left-right gluonic penguins, $Q_5$ and $Q_6$.  Table 6 compares
the results of the PQS method versus those of the PQN method.  The
results are for the subtracted $K \to \pi$ matrix elements, where
the (large) subtraction is performed using $K\to 0$.  For details
on how this subtraction is performed, see \cite{blum}. The end
result of this subtraction in the case of full QCD (no quenching)
is just $\alpha_1^{N=3}$, which is the physical LO LEC that
contributes to $K\to\pi\pi$.  In this case, the two methods, PQS
and PQN, are procedurally the same, and they therefore give the
same answer.

\begin{table}[htbp]
\caption{The leading order LEC's as determined in PQChPT from
$K\to\pi$ after using the $K\to0$ subtraction described in
\cite{blum} are presented.  They are compared for the PQS and PQN
methods in the case of the left-right gluonic penguins.  The two
methods agree for the full QCD case.  For the $N=3$, $m_{sea}\neq
m_{val}$ case there is a logarithmic contamination for the case of
PQN.  That is, there is an extra term appearing at leading order
that must be accounted for in fits used to obtain the subtraction
coefficient from $K\to0$.  For $N=2$, the LEC's are not those of
the full theory, and additional terms appear for the PQN case. For
$N=0$, the quenched case, there are also additional terms that
contribute in the PQN case.  See \cite{goltthree,goltfour} for the
derivations of these results and the values of the logarithmic
corrections abbreviated here.}
\begin{center}
\begin{tabular}{|c|c|}
  \hline
  &  $N=3$, $m_{sea}= m_{val}$ (Full QCD) \\
  \hline
  PQS & PQN \\
  \hline
    & \\
  $\alpha^{N=3}_1$ &  $\alpha_1^{N=3}$ \\
     &  \\
  \hline
  \hline
  &  $N=3$, $m_{sea}\neq m_{val}$ \\
  \hline
  PQS & PQN \\
  \hline
    & \\
  $\alpha^{N=3}_1$ &  $\alpha_1^{N=3} + \alpha^{N=3}_{(8,8)}(log$ $terms)$ \\
     &  \\
  \hline
  \hline
   & $N=2$ \\
   \hline
  PQS & PQN \\
  \hline
    &  \\
  $\alpha^{N=2}_1$ & $\frac{3}{2}\alpha_1^{N=2}+\frac{1}{(4\pi)^2}(\beta^{(8,8)}_1+\frac12 \beta^{(8,8)}_2)
  + \alpha^{(8,8)}_{N=2}(log$ $terms)$ \\
   &  \\
  \hline
  \hline
  &  $N=0$  \\
  \hline
   PQS & PQN \\
  \hline
    &  \\
  $\alpha_1^{N=0}$ & $\frac{1}{2}\alpha_1^{N=0}-\frac{1}{(4\pi)^2}(\beta^{NS}_1+\frac12 \beta^{NS}_2)
  + \alpha^{NS}_Q(log$ $terms)$ \\
   &  \\
   \hline

\end{tabular}
\end{center}
\end{table}

When $N=3$, but $m_{sea}\neq m_{val}$, the LEC's in the amplitudes
are still those of the full theory, but an additional LEC, the
leading order (8,8) electroweak penguin LEC, $\alpha_{88}$,
contributes in the PQN case to $K\to 0$ multiplied by some
logarithmic terms \cite{goltthree}.  Thus, a subtraction that is
performed without taking this into account has a contamination.
That is, there is an extra term appearing at leading order that
must be accounted for in fits used to obtain the subtraction
coefficient from $K\to0$.  Looking in Table 6 at the PQS result,
we see that this method is simpler. At NLO the difference is even
more severe, so that PQS is the only method shown to be feasible.
In this case the difference in practice between the two methods is
whether one uses the sea mass or the valence mass in the
propagator of the loop in Fig 3c. (See the preceding section for
the correspondence between Fig 3c and the traditional form of the
eye-diagrams in Fig 2 for the various operators.)

When $N=2$ the LEC's are no longer those of the full theory, and
an ambiguity results.  In this case the calculations differ in
that for the loop of Fig 3c for PQS one uses 2 flavors with the
dynamical mass, while for PQN one uses 3 flavors with the valence
masses.  Here, $\alpha_{88}^{N=2}$ appears multiplied by
logarithmic terms, and these must be removed in the fits to
$K\to0$ before the subtraction can be performed.  Notice also the
presence of the $\beta$ terms in linear combination with the
$\alpha_1^{N=2}$ term.  These $\beta$ terms always appear in the
same linear combination with $\alpha_1^{N=2}$, including in the
expression for $K\to\pi\pi$.  Thus, it is not obvious whether they
represent a correction to the $\alpha_1^{N=2}$ term or a
contamination.  Clearly, it will be important to compare the
results of both methods.  Note also that the $\beta$ terms that
appear with $\alpha_1^{N=2}$ have a scale dependence proportional
to $\alpha_{88}^{N=2}$, and that if $\alpha_{88}^{N=2}$ is not so
far from $\alpha_{88}^{N=0}$, as determined in \cite{blum}, then
this scale dependence would be large.  It would then be necessary
to include the partially quenched chiral logs proportional to
$\alpha_{88}^{N=2}$ in $K\to\pi\pi$ in order to cancel the scale
dependence and obtain a consistent answer.  It would be extremely
useful to have (at least so long as an $N=3$ calculation is not
available) a study of the subtracted LO constants for the PQS and
PQN methods as a function of $N$, so that one could try to
extrapolate each result to $N=3$, and compare the two.

Note that the $\beta$ terms are related to the (8,8) LEC's, $c_i$
in the terminology of this paper.  The correspondence to our
notation is

\begin{eqnarray}\label{11}
\beta_1^{(8,8)}=(4\pi)^2 2c_3, \nonumber \\
\beta_2^{(8,8)}=(4\pi)^2 2c_1,  \nonumber \\
\beta_3^{(8,8)}=(4\pi)^2 2c_4.
\end{eqnarray}

Finally, when $N=0$ the theory is completely quenched.  This
corresponds to ignoring all contractions of the kind in Fig 3c for
the PQS method and keeping them with the valence quarks for PQN.
In the quenched case, the additional LEC's for PQN, the
$\alpha^{NS}_Q$ and $\beta^{NS}_i$, are coefficients of
non-singlet operators, but they have no relation to the (8,8)
electroweak penguins.  Golterman and Pallante \cite{goltthree}
provide a possible recipe for determining $\alpha^{NS}_Q$ on the
lattice. Analogous to the partially quenched case, if
$\alpha^{NS}_Q$ is large, it would imply a large scale dependence
on the subtracted combination of LEC's in the PQN method, which
would have to be cancelled by the quenched logs proportional to
$\alpha^{NS}_Q$ in $K\to\pi\pi$ in order to obtain a consistent,
scale independent answer.  There are indications from the large
$N_c$ ($N_c$ is the number of colors) approximation that the LEC,
$\alpha^{NS}_Q$, is indeed large compared to $\alpha_1^{N=0}$
\cite{goltfive}. Again, it is not clear if the non-singlet terms
represent a correction or a contamination, and results for both
methods should be compared as part of an extrapolation in $N$.

\section{(Partially) Quenched (8,8)'s to NLO}

In this section we present the results for the partially quenched
$K\to\pi$ and $K\to0$ amplitudes needed to construct the $K
\to\pi\pi$ amplitudes to NLO for the (8,8)'s.  The power divergent
subtraction is discussed for the $\Delta I=1/2$, $K\to\pi$
amplitude.  Formulas are presented for $K\to0$ and $K\to\pi$ for
nondegenerate quark masses, as well as $K\to\pi$ for degenerate
quark masses.  It is demonstrated that $K\to \pi$ with degenerate
masses is sufficient to construct $K\to\pi\pi$ to NLO in the
partially quenched theory, while $K\to\pi$ with non-degenerate
quark masses gives additional redundancy in determining the NLO
LEC's.

We show that in the case of $K\to\pi$ with degenerate quark mass,
the $N=0$ limit of our expressions produces the quenched result,
which will be useful for fits to already existing lattice data. It
is important to notice that not all LEC's needed for NLO
$K\to\pi\pi$ can be determined from the quenched $K\to\pi$ data,
since $c^r_6$, which is needed in the physical $K\to\pi\pi$
expressions, does not appear in the quenched $K\to\pi$ formulas.
One can see the scale dependence of the LEC's from the formula,

\begin{equation}
c^r_i(\mu_2)=c^r_i(\mu_1)+\frac{2\alpha_{88}\eta_i}{(4\pi
f)^2}\ln{\frac{\mu_1}{\mu_2}},
\end{equation}

\noindent which can be obtained from Eq (19), the definition of
the renormalized LEC's.  The coefficients, $\eta_i$, are given in
Table 3.  Since the scale dependence of the $c^r_6$ coefficient in
the physical $K\to\pi\pi$ amplitude is significant (where it is
needed to cancel the corresponding scale dependence in the NLO log
terms), it is crucial to do dynamical simulations of the (8,8)
$K\to\pi$ amplitudes in order to bring under control the
systematic errors due to the chiral expansion.

\subsection{Partially Quenched (8,8)'s with nondegenerate quark masses}

The LEC's needed to construct the $K \rightarrow \pi \pi$, $\Delta
I=1/2$ and 3/2 (8,8)'s can be obtained from the $K \rightarrow
\pi$ amplitudes with energy insertion and $m_s \neq m_d=m_u$.  The
(8,8) $K \rightarrow \pi \pi$ counterterm contributions for both
the $ \Delta I=3/2$ and 1/2 amplitudes are given by

\bea \langle \pi^{+}\pi^{-}|{\cal
O}^{(8,8),(3/2)}|K^{0}\rangle_{ct} &=& -\frac{4i\alpha_{88}}{f_K
f^{2}_\pi}+\frac{4i}{f_K
f^2_\pi}[(-c^r_2-c^r_3-2c^r_4-2c^r_5-4c^r_6)m^2_K \nonumber
\\ && -(-c^r_1-c^r_2+4c^r_4+4c^r_5+2c^r_6)m^2_\pi],
\eea

\bea \langle \pi^{+}\pi^{-}|{\cal
O}^{(8,8),(1/2)}|K^{0}\rangle_{ct}& =& -\frac{8i\alpha_{88}}{f_K
f^{2}_\pi}-\frac{4i}{f_K
f^2_\pi}[(-c^r_1-c^r_2+4c^r_4+4c^r_5+8c^r_6)m^2_K \nonumber \\
&&+(-c^r_1+c^r_2+2c^r_3+8c^r_4+8c^r_5+4c^r_6)m^2_\pi]. \eea

These are the expressions in the full theory and were given by
\cite{cirig}, where they showed that one can obtain the necessary
linear combinations of LEC's from $K \rightarrow \pi$ with
momentum, $\Delta I=1/2$, 3/2.  We demonstrate this holds also for
the partially quenched case (without the need for 3-momentum
insertion, as explained in Section 5). In Eqs (42), (43) as well
as all the following amplitudes, we include only the tree level
weak counterterm contributions.  For clarity, the logarithmic
terms and the Gasser-Leutwyler $L_i$ counterterms have been
omitted from this section, but are included in Appendix D.

The $K \rightarrow \pi$ counterterm amplitudes are given by

\bea \langle \pi^{+}|{\cal O}^{(8,8),(3/2)}|K^{+}\rangle_{ct}& =&
\frac{4\alpha_{88}}{f^{2}}+\frac{4}{f^2}[2(c^r_4+c^r_5)m^2_K +2(c^r_4+c^r_5)m^2_\pi\nonumber \\
&&-(c^r_1+c^r_2)m_K m_\pi+2c^r_6Nm^2_{SS}], \eea

\bea \langle \pi^{+}|{\cal O}^{(8,8),(1/2)}|K^{+}\rangle_{ct}& =&
\frac{8\alpha_{88}}{f^{2}}+\frac{4}{f^2}[(6c^r_4+4c^r_5)m^2_K +4(c^r_4+c^r_5)m^2_\pi\nonumber \\
&&-(c^r_1-c^r_2-2c^r_3)m_K m_\pi+4c^r_6Nm^2_{SS}]. \eea

At this point, a practical issue in the extraction of the LEC's
should be mentioned.  There is a power divergence in the NLO
coefficient $c^r_4$ due to mixing with unphysical lower
dimensional operators that must be removed if one is to have any
hope of numerically extracting any of the LEC's. This is a problem
for $K\to\pi$, $\Delta I=1/2$, but not $K\to\pi$, $\Delta I=3/2$,
since the combination $c^r_4+c^r_5$ is finite in the continuum
limit. For the $\Delta I=1/2$ amplitude the power subtraction
method of RBC \cite{blum} can be used, and this requires the
$K\to0$ amplitude.  At NLO, this amplitude is

\begin{eqnarray}\label{15}
\langle 0|{\cal O}^{(8,8)}|K^{0}\rangle &=&
\frac{4i\alpha_{88}}{f}[2A_0(m^2_K)-A_0(m^2_\pi)-A_0(m^2_{33})
\nonumber \\ && +NA_0(m^2_{sS})-NA_0(m^2_{uS})]
 -
\frac{8i}{f}c^r_4(m^2_K-m^2_\pi),
\end{eqnarray}

\noindent where $A_0(m^2)$ is defined in Appendix A.  We also
mention that the (8,8), $K\to0$ calculation has an eye-diagram,
and one must make a decision whether to keep the valence quarks in
the eye contractions or not. The above formula, Eq (46),
corresponds to keeping the valence quarks in the eye contraction.
If one neglects the type of contraction associated with Fig 3c,
then one obtains

\begin{eqnarray}\label{15}
\langle 0|{\cal O}^{(8,8)}|K^{0}\rangle &=&
\frac{4i\alpha_{88}}{f}[NA_0(m^2_{sS})-NA_0(m^2_{uS})]
 -
\frac{8i}{f}c^r_4(m^2_K-m^2_\pi). \nonumber \\ &&
\end{eqnarray}

\noindent Unlike the case of $Q_6$, the chiral perturbation theory
is not substantially changed, and one can use either method, as
long as one is consistent.  The $K\to0$ subtraction works as
follows. We make use of the subtraction operator introduced in
Section 4,

\begin{equation}
\Theta^{(3,\overline{3})}\equiv \overline{s}(1-\gamma_5)d=
\alpha^{(3,\overline{3})}\textrm{Tr}(\lambda_6\Sigma)
\end{equation}

\noindent to lowest order in chiral perturbation theory.  The mass
dependence of the above quark bilinear operator,
$\Theta^{(3,\overline{3})}$, is the same as that of the power
divergent part of the four-quark operators, so one can use the
matrix elements of this bilinear operator to subtract out power
divergences to all orders in ChPT. In order to perform the
subtraction at NLO for the (8,8)'s we need the following leading
order expressions of the $\Theta^{(3,\overline{3})}$ amplitudes,

\begin{equation}
\langle \pi^{+}|\Theta^{(3,\overline{3})}|K^{+}\rangle =
\frac{-2}{f^2}\alpha^{(3,\overline{3})},
\end{equation}

\begin{equation}
\langle 0|\Theta^{(3,\overline{3})}|K^{0}\rangle =
\frac{2i}{f}\alpha^{(3,\overline{3})}.
\end{equation}

When we take the ratio of $\langle 0|{\cal
O}^{(8,8)}|K^{0}\rangle$ to $\langle
0|\Theta^{(3,\overline{3})}|K^{0}\rangle$ we get

\begin{equation}
\frac{\langle 0|{\cal O}^{(8,8)}|K^{0}\rangle}{\langle
0|\Theta^{(3,\overline{3})}|K^{0}\rangle} =
-4\frac{c^r_4}{\alpha^{(3,\overline{3})}}(m^2_K-m^2_\pi)+2\frac{\alpha_{88}}{\alpha^{(3,\overline{3})}}(logs)+...
\end{equation}

\noindent where we have omitted terms of higher order in the
chiral expansion.  Note, however, that all higher order terms
proportional to $c^r_4$ cancel in the ratio.  Fitting to this
expression allows one to obtain $c^r_4/\alpha^{(3,\overline{3})}$,
which one can then use in the subtraction of the power divergences
of $K\to\pi$.  Notice that $c^r_4$ has a scale dependence that
must cancel the scale dependence of the $\alpha_{88}$ log term in
$K \to0$.  Thus, one must ensure the value of $\mu$ in a chiral
fit to $K\to\pi$ is the same as the value of $\mu$ used in the
power subtraction.  After the subtraction, the following
expression no longer has power divergences.

\begin{eqnarray}\label{15}
\langle \pi^+|{\cal O}^{(8,8),(1/2)}|K^{+}\rangle &+& 4\frac{c^r_4
m^2_K}{\alpha^{(3,\overline{3})}}\langle
\pi^+|\Theta^{(3,\overline{3})}|K^{+}\rangle =
\frac{8\alpha_{88}}{f^2}(1+logs)+\frac{4}{f^2}[4(c^r_4+c^r_5)m^2_K
\nonumber \\ && +4(c^r_4+c^r_5)m^2_\pi-(c^r_1-c^r_2-2c^r_3)m_K
m_\pi+4Nc^r_6 m^2_{SS}].
\end{eqnarray}

\noindent Here, $c^r_4$ appears only in the linear combination
$c^r_4+c^r_5$, which does not contain power divergences.  Thus,
the $K\to0$ subtraction has removed the power divergences from the
$\Delta I=1/2$, $K\to\pi$ expression, including all of the higher
order power divergent contributions, an important point, since the
subtraction does not require (PQ)ChPT for its implementation.  The
expression, Eq (52), is the one which should be fitted for the NLO
LEC's.  Thus, fitting to (44) and the power subtracted amplitude,
(52), one can obtain all of the linear combinations needed for
$K\to\pi\pi$ at NLO.

In principle, one can obtain $\alpha_{88}$ from either leading
order term.  In practice, it is safer to use the 3/2 amplitude
since the 1/2 amplitude could receive some residual chiral
symmetry breaking contribution unless one uses a discretization
that has exact chiral symmetry.  The $\Delta I=3/2$ expression
does not involve power divergent subtractions, and is, therefore,
the best way to get the leading order coefficient.  One can get
$c^r_6$ from the term that depends on the sea meson mass. From
fits to the other mass combinations, one obtains $c^r_4+c^r_5$,
$c^r_1+c^r_2$, and $c^r_1-c^r_2-2c^r_3$. Along with $\alpha_{88}$,
the four linear combinations: $[c^r_1+c^r_2, c^r_1-c^r_2-2c^r_3,
c^r_4+c^r_5,c^r_6]$ are sufficient to determine $K \rightarrow \pi
\pi$ at the physical kinematics, as one can verify with some
simple algebra from Eqs (42) and (43). When $N=3$, the values of
the LEC's determined from PQChPT are the same as in the full
theory.  We point out in the next subsection that one can get all
of the needed information to construct the EWP matrix element for
$K\to\pi\pi$ to NLO even with $K\to\pi$ using degenerate valence
quark masses, along with $K\to0$ to perform the power subtraction
in the $\Delta I=1/2$ case.  The nondegenerate case remains
useful, however, in that it provides additional redundancy in
determining the NLO LEC's.

Note that for the cases of physical $K \rightarrow \pi \pi$ (8,8)
amplitudes (42),(43), and (58) for the corresponding (8,1)'s, the
pseudoscalar decay constants and masses are the physical
(renormalized to one-loop order) ones. For all other amplitudes
given in this paper except $K \rightarrow \pi \pi$ at physical
kinematics, the formulas are in terms of the bare constants.  The
distinction between bare and renormalized constants is made only
in tree-level amplitudes, since making this distinction in the NLO
expressions introduces corrections at higher order (NNLO) than is
considered here.

The logarithmic and Gasser-Leutwyler counterterm contributions to
the amplitudes in this section are given in Appendix D.

\subsection{Partially Quenched (8,8)'s with degenerate quark masses for $K\to\pi$}

For the case of degenerate quark masses, Eqs (44) and (52) become

\bea \langle \pi^{+}|{\cal O}^{(8,8),(3/2)}|K^{+}\rangle_{ct}& =&
\frac{4\alpha_{88}}{f^{2}}+\frac{4}{f^2}[(-c^r_1-c^r_2+4c^r_4+4c^r_5)m^2
+2c^r_6Nm^2_{SS}], \nonumber \\ && \eea

\bea \langle \pi^{+}|{\cal
O}^{(8,8),(1/2)}_{sub}|K^{+}\rangle_{ct}& =&
\frac{8\alpha_{88}}{f^{2}}+\frac{4}{f^2}[(-c^r_1+c^r_2+2c^r_3+8c^r_4+8c^r_5)m^2
+4c^r_6Nm^2_{SS}]. \nonumber \\ && \eea

\noindent where the logarithmic parts of the above expressions are
given in Appendix D.  The $K\to0$ subtraction is performed exactly
as in the non-degenerate case, yielding the above result, Eq (54).
Again, simple algebra will verify that the above linear
combinations of LEC's are sufficient to determine the LEC
combinations in Eqs (42) and (43) for the physical (8,8)
$K\to\pi\pi$ amplitudes to NLO.  For example, if one subtracts the
$m^2$ coefficient in Eq (54) from the $m^2$ coefficient in Eq
(53), one gets the same linear combination as the first four terms
in Eq (42). If one wants to obtain all the needed information in
the full theory, one must do an $N=3$ simulation, varying the sea
quark mass with respect to the valence quark mass in order to
determine $c^r_6$.  That is, one must still vary the sea quark
mass independently of the valence quark mass.

\subsection{Quenched (8,8)'s with degenerate quark masses for $K\to\pi$}

One can obtain results in the quenched theory by taking the $N=0$
limit of the $K\to0$ and degenerate $K\to\pi$ formulas. One then
obtains, for $K\to0$,

\begin{eqnarray}\label{15}
\langle 0|{\cal O}^{(8,8)}|K^{0}\rangle &=&
\frac{4i\alpha_{88}}{f}[2A_0(m^2_K)-A_0(m^2_\pi)-A_0(m^2_{33})]
 -\frac{8i}{f}c^r_4(m^2_K-m^2_\pi). \nonumber \\ &&
\end{eqnarray}

\noindent In the quenched theory, the scale dependence of $c^r_4$
vanishes, as one can verify from Table 3.  Therefore, the scale
dependence of the logarithms in quenched $K\to0$ must also vanish;
the fact that it does can be seen from Eq (55).  The subtraction
is performed the same way as in the partially quenched theory, and
the expressions for $K\to\pi$ are

\bea \langle \pi^{+}|{\cal O}^{(8,8),(3/2)}|K^{+}\rangle_{ct}& =&
\frac{4\alpha_{88}}{f^{2}}\biggl[1-\frac{2}{16\pi^2f^2}\left(m^2\ln\frac{m^2}{\mu^2}+m^2\right)\biggr]\nonumber
\\ && +\frac{4m^2}{f^2}\left(\frac{-16\alpha_{88}}{f^2}L^Q_5-c^r_1-c^r_2+4c^r_4+4c^r_5\right),
\nonumber \\ && \eea

\bea \langle \pi^{+}|{\cal
O}^{(8,8),(1/2)}_{sub}|K^{+}\rangle_{ct}& =&
\frac{8\alpha_{88}}{f^{2}}\biggl[1+\frac{1}{16\pi^2f^2}\left(m^2\ln\frac{m^2}{\mu^2}+m^2\right)\biggr]\nonumber
\\ && +\frac{4m^2}{f^2}\left(\frac{-32\alpha_{88}}{f^2}L^Q_5-c^r_1+c^r_2+2c^r_3+8c^r_4+8c^r_5\right).
\nonumber \\ && \eea

\noindent Here, $L^Q_5$ is the quenched Gasser-Leutwyler
coefficient that appears in $f_\pi$.  Notice that $c^r_6$ does not
appear in the quenched theory, though it does appear in
$K\to\pi\pi$, where, as one can see from Table 3 it has a
non-vanishing scale dependence.  This dependence on the chiral
scale leads to a large uncertainty in the NLO $K\to\pi\pi$
amplitudes, since the scale dependence of the LEC's must cancel
against those of the chiral logarithms.  Thus, it is essential to
compute $K\to\pi$ with dynamical quarks in order to reduce the
uncertainty due to the chiral expansion.

\section{Partially Quenched (8,1)'s to NLO}

The amplitudes necessary to construct the physical $K \rightarrow
\pi \pi$ matrix elements are $K \rightarrow 0$; $K \rightarrow
\pi$, $m_s \neq m_d=m_u$; and $K \rightarrow \pi \pi$ at the
unphysical kinematics points of UKX, of which UK1 and UK2 are
special cases. The counterterm part of the physical (8,1)
amplitude is

\bea \langle \pi^{+}\pi^{-}|{\cal O}^{(8,1)}|K^{0}\rangle_{ct}& =&
\frac{4i\alpha_{1}}{f_K
f^{2}_\pi}(m^2_K-m^2_\pi)_{(1-loop)}+\frac{8i}{f_K
f^2_\pi}(m^2_K-m^2_\pi)\nonumber \\ &&
\times\biggl[(e^r_{10}-2e^r_{13}+2e^r_{14}+e^r_{15})m^2_K
+(-2e^r_1+2e^r_{10}\biggr. \nonumber
\\ && \biggl.+e^r_{11}+4e^r_{13}+e^r_{14}-4e^r_2-2e^r_3-4e^r_{35}+8e^r_{39})m^2_\pi
\nonumber \\ && +\frac{8\alpha_2}{f^2}[2m^2_K L_4 +(-4L_4-L_5+8L_6+4L_8)m^2_\pi] \biggr]. \nonumber \\
\eea

This differs from our previous expression \cite{laiho}, Eq. (35)
in the appearance of a new LEC, $e_{14}^r$, and in the inclusion
of the Gasser-Leutwyler coefficients of the amplitude previously
given separately as part of the log terms in D10 of \cite{laiho}.
The Gasser-Leutwyler coefficients are included here with the rest
of the LEC's for clarity.  In the full theory the operator
corresponding to this LEC can be absorbed into the other operators
${\cal O}^{(8,1)}_{10}$, ${\cal O}^{(8,1)}_{11}$, ${\cal
O}^{(8,1)}_{12}$ and ${\cal O}^{(8,1)}_{13}$ via the
Cayley-Hamilton theorem, as discussed in Section 4.  Since this is
no longer true in the partially quenched theory, one must obtain
the constant separately.  Therefore, it is left explicit in the
physical amplitude.

For $K^0 \rightarrow 0$, we have

\begin{eqnarray}\label{15}
\langle 0|{\cal O}^{(8,1)}|K^{0}\rangle_{ct} &=&
\frac{4i\alpha_{2}}{f}(m^{2}_{K}-m^{2}_{\pi})
 +
\frac{8i}{f}(m^2_K-m^2_\pi)[(2e^r_1-2e^r_5)m^2_K\nonumber \\
&& +e^r_2 Nm^2_{SS}].
\end{eqnarray}

The expression for $K\to\pi$ is

\begin{eqnarray}\label{19}
\langle \pi^{+}|{\cal O}^{(8,1)}|K^{+}\rangle_{ct} &=&
\frac{4}{f^{2}}\alpha_{1}m_{K}m_{\pi} -
\frac{4}{f^{2}}\alpha_{2}m^{2}_{K}  -
\frac{8}{f^{2}}[2(e^{r}_{1}-e^r_5)m^{4}_{K}\nonumber \\
&&-2(e^{r}_{10}-e^{r}_{35}) m^{3}_{K}m_{\pi}
 + (2e^{r}_{3}+2e^{r}_{5}-8e^r_{39})m^{2}_{K}m^2_\pi \nonumber \\ && +
(2e^{r}_{35}-e^{r}_{11})m_K m^3_\pi +Ne^{r}_{2}m^2_K
m^2_{SS}-Ne^r_{14}m_K m_\pi m^2_{SS}]. \nonumber \\
\end{eqnarray}

The (8,1) amplitudes have power divergent parts that must be
subtracted, just as in the case of the $\Delta I=1/2$, (8,8)'s.
The subtraction is performed in the same way, but in this case the
power divergent coefficient, $\alpha_2$, is present already at
leading order.  Thus, we must consider the effects of the power
subtraction at NLO if we are interested in obtaining the LEC's to
this order.  As we will see, the effect is to modify the NLO,
(8,1) LEC's by adding terms proportional to the Gasser-Leutwyler
coefficients.  The LEC's modified in this way are just those that
have a scale dependent part proportional to $\alpha_2$, and the
new constants so obtained are given in Table 5.  The power
subtraction eliminates the tadpole contributions from the
amplitudes, and the new combinations of LEC's in Table 5 are free
of power divergences, and can be obtained in numerical fits to
lattice data.

The ratio of the $K\to0$ amplitude to the
$\Theta^{(3,\overline{3})}$ $K\to0$ amplitude to NLO in PQChPT is

\begin{eqnarray}
\frac{\langle 0|{\cal O}^{(8,1)}|K^{0}\rangle}{\langle
0|\Theta^{(3,\overline{3})}|K^{0}\rangle} &=&
2\frac{\alpha_2}{\alpha^{(3,\overline{3})}}(m^2_K-m^2_\pi)+2\frac{\alpha_1}{\alpha^{(3,\overline{3})}}(logs)
+\frac{4}{\alpha^{(3,\overline{3})}}(m^2_K-m^2_\pi) \nonumber
\\ && \times \left[2\left(e^r_1-\frac{8\alpha_2}{f^2}L^r_8
-e^r_5\right)m^2_K+\left(e^r_2-\frac{16\alpha_2}{f^2}L^r_6\right)Nm^2_{SS}\right].\nonumber
\\ &&
\end{eqnarray}

In this case, $\alpha_2$ appears multiplied by $m^2_K-m^2_\pi$
(these are the tree-level masses, directly proportional to
$m_s-m_d$), but all higher order logarithmic terms proportional to
$\alpha_2$ are subtracted in the ratio. Also, the NLO terms from
the $\Theta^{(3,\overline{3})}$ operator appear in just the
combinations given in Table 5.  As we will show, the effect of the
subtraction on $K\to\pi$ and $K\to\pi\pi$ is to eliminate the
$\alpha_2$ term, including all higher order corrections
proportional to $\alpha_2$, and the NLO LEC's are modified to the
values in Table 5, just as in the ratio for $K\to0$. After the
subtraction of $K\to\pi$, at NLO one is left with

\begin{eqnarray}\label{15}
\langle \pi^+|{\cal O}^{(8,1)}|K^{+}\rangle &-& 2\frac{\alpha_2
m^2_K}{\alpha^{(3,\overline{3})}}\langle
\pi^+|\Theta^{(3,\overline{3})}|K^{+}\rangle =
\frac{4\alpha_1}{f^2}m_K m_\pi(1+logs)-\frac{8}{f^2}[2(e^{r}_{1,rot}-e^r_{5,rot})m^{4}_{K}\nonumber \\
&&-2(e^{r}_{10,rot}-e^{r}_{35}) m^{3}_{K}m_{\pi}
 + (2e^{r}_{3,rot}+2e^{r}_{5,rot}-8e^r_{39})m^{2}_{K}m^2_\pi \nonumber \\ && +
(2e^{r}_{35}-e^{r}_{11})m_K m^3_\pi +Ne^{r}_{2,rot}m^2_K
m^2_{SS}-Ne^r_{14}m_K m_\pi m^2_{SS}].
\end{eqnarray}

\noindent where we have indicated the coefficients that have
undergone a chiral rotation to the form of Table 5 with a
subscript, for brevity.  As expected, the dependence on $\alpha_2$
vanishes.

The amplitudes in the partially quenched theory for $K\to\pi\pi$
at UK1 and UK2 are

\begin{eqnarray}\label{20}
\langle \pi^{+} \pi^{-}|{\cal O}^{(8,1)}|K^{0}\rangle_{ct}& =& 8i
\frac{\alpha_{1}}{f^{3}}m^{2}
  + 8i\frac{m^{2}}{f^{3}}[(4e^{r}_{10}+ 2e^{r}_{11}+4e^{r}_{15}
 -4e^{r}_{35})m^2\nonumber \\ && +2Nm^2_{SS}e^r_{14}],
\end{eqnarray}

\noindent for $ K \rightarrow \pi \pi, m_{K}=m_{\pi}=m $ (UK1),
and

\begin{eqnarray}\label{22}
\langle \pi^{+} \pi^{-}|{\cal O}^{(8,1)}|K^{0}\rangle_{ct} &=&
4i\frac{\alpha_{1}}{f^{3}}(m^{2}_{K}-m^2_\pi)
 + \frac{3i}{2}\frac{m^{2}_{K}}{f^{3}}\nonumber \\
 && \times[(-2e^{r}_{1}+
6e^{r}_{10}+e^{r}_{11}-4e^{r}_{13}
 + 4e^{r}_{15}-4e^{r}_{2} \nonumber \\ &&
 -2e^{r}_{3}-4e^{r}_{35}+8e^{r}_{39})m^2_K+4e^r_{14}Nm^2_{SS}]\nonumber \\ &&
 +12i \frac{\alpha_2}{f^5} m^4_K  \left( 4L_4 -L_5 +8L_6
+4L_8\right),
\end{eqnarray}

\noindent for $ K \rightarrow \pi \pi,
m_{K(1-loop)}=2m_{\pi(1-loop)}$ (UK2).

In the partially quenched case there are, in general, additional
complications in the calculation of the $ \Delta I=1/2$, $K
\rightarrow \pi \pi$ amplitudes due to threshold divergences
leading to enhanced finite volume effects \cite{colang, lintwo}
which require special care. We present the logarithmic terms for
the infinite volume Minkowski space amplitudes at these kinematics
in Appendix E. The threshold divergences are imaginary and vanish
at NLO when the sea meson mass becomes equal to the pion mass (or
in terms of quarks, $m_{sea}=m_u=m_d$) for any $N \geq 1$, see Eqs
(E3, E5). In infinite volume Euclidean space one might expect the
imaginary part should vanish since ${\cal M}_{Euclid}=1/2({\cal
M}|_{in}-{\cal M}|_{out})$, but in lattice calculations the
imaginary part shows up in the form of enhanced finite volume
effects.  In finite volume Euclidean correlation functions, unless
the sea masses are chosen as stated ($m_{sea}=m_u=m_d$), these
enhanced finite volume effects are present, and they diverge as a
power of the lattice volume. See \cite{bernfour,lintwo} for
relevant calculations and discussions.\footnote{Ref
\cite{linthree} states that in partially quenched lattice
calculations, unless the unphysical degrees of freedom are above
the two pion threshold, one may have serious problems with the
enhanced finite volume effects. Again, see the note added in
revision.}

As pointed out in Section 5, there exists a set of kinematics for
$K\to\pi\pi$ where the kaon and both pions are at rest, bypassing
the Maiani-Testa theorem on the lattice.  The quark masses ($m_s$
and $m_u=m_d$) can be varied independently, where the weak
operator inserts/removes energy to enforce 4-momentum
conservation.  We call this set of kinematics UKX, of which UK1
and UK2 are special cases.  As pointed out by \cite{goltsix},
there is a subtlety involved in calculating the UK1 kinematics,
and the $\epsilon$ prescription must be applied.  One must take
the $\epsilon \rightarrow 0$ limit only after setting $m_K=m_\pi$.
Given below is the most general expression for the LEC
contribution to UKX,

\begin{eqnarray}\label{19}
\langle \pi^{+}\pi^{-}|{\cal O}^{(8,1)}|K^{0}\rangle_{ct} &=&
\frac{4i}{f^{3}}\alpha_{1}m_{\pi}(m_K+m_{\pi}) +
\frac{8i}{3f^{3}}\alpha_{2}(m^2_K-m^2_\pi)\frac{3m_\pi(2m_\pi-m_K)+i\epsilon}{4m_\pi(m_K-m_\pi)-i\epsilon}\nonumber
\\ && +\frac{16i(m_K^2-m^2_\pi)}{3f^3[4m_\pi(m_K-m_\pi)-i\epsilon]}
\left\{[(4m^2_\pi-4m_K m_\pi+i\epsilon)(2m^2_K+3m^2_\pi) \right.
\nonumber \\ &&  +2m^2_Km_\pi(m_K+2m_\pi)]e_1^r +
m^2_{SS}[3(2N+8)m^2_\pi
\nonumber \\ && \left. -3(N+8)m_K m_\pi +(N+6)i\epsilon]e^r_2
-2m^2_K(6m^2_\pi-3m_Km_\pi+i\epsilon)e^r_5 \right\} \nonumber\\
&& +\frac{8i}{f^{3}}[(2e^r_{10}-2e^{r}_{35}) m^{3}_{K}m_{\pi}
\nonumber \\ &&
 + (-2e^{r}_{3}-4e^r_{13}+2e^r_{15}+2e^r_{35}+8e^r_{39})m^{2}_{K}m^2_\pi
 \nonumber \\ && +(e^r_{11}+2e^{r}_{15})m_K m^3_\pi \nonumber \\ &&
 +(2e^r_3+2e^r_{10}+e^r_{11}+4e^r_{13}-4e^r_{35}-8e^r_{39})m^4_\pi
 +Ne^r_{14}m_K m_\pi m^2_{SS}]. \nonumber \\
\end{eqnarray}

When $m_K=m_\pi$, and the limit $\epsilon \rightarrow 0$ is taken
in Eq (65), we recover the special case of UK1, given by Eq (63).
One can only use UKX within the range $m_K>m_\pi$ \cite{linthree},
though we show that all LEC's can still be determined.  The LEC
contribution to UKX reduces to Eq (66) at the special kinematics
$m_K>m_\pi$ and $m_{sea}=m_u=m_d$ ($m_{SS}=m_\pi$), where the
imaginary threshold divergences vanish.

\begin{eqnarray}\label{19}
\langle \pi^{+}\pi^{-}|{\cal O}^{(8,1)}|K^{0}\rangle_{ct} &=&
\frac{4i}{f^{3}}\alpha_{1}m_{\pi}(m_K+m_{\pi}) +
\frac{2i}{f^{3}}\alpha_{2}(m_{K}+m_\pi)(2m_\pi-m_K)\nonumber \\ &&
+\frac{8i}{f^{3}}[(-e^{r}_{1}+e^r_5)m^{4}_{K}+(e^{r}_{1}-e^r_5+2e^r_{10}-2e^{r}_{35})
m^{3}_{K}m_{\pi} \nonumber \\ &&
 + ((-4-N/2)e^r_2-2e^{r}_{3}-2e^{r}_{5}-4e^r_{13}+2e^r_{15}+2e^r_{35}+8e^r_{39})m^{2}_{K}m^2_\pi
 \nonumber \\ && +((N/2)e^r_2+e^r_{11}+Ne^{r}_{14}+2e^{r}_{15})m_K m^3_\pi \nonumber \\ &&
 +(2e^r_1+(4+N)e^{r}_{2}+2e^r_3+2e^r_{10}+e^r_{11}+4e^r_{13}+Ne^r_{14}-4e^r_{35}-8e^r_{39})m^4_\pi]. \nonumber \\
\end{eqnarray}

\noindent The logarithmic part of this expression is given by Eq
(E10).

The power divergent subtraction must also be performed on $K\to
\pi\pi$ amplitudes, and this requires the computation of the
matrix element, $\langle
\pi^+\pi^-|\Theta^{(3,\overline{3})}|K^{0}\rangle$.  The
subtraction to be performed is

\begin{equation}
\langle \pi^+\pi^-|{\cal
O}^{(8,1)}_{sub}|K^{0}\rangle\equiv\langle \pi^+\pi^-|{\cal
O}^{(8,1)}|K^{0}\rangle - 2\frac{\alpha_2
}{\alpha^{(3,\overline{3})}}(m^2_K-m^2_\pi)\langle
\pi^+\pi^-|\Theta^{(3,\overline{3})}|K^{0}\rangle, \nonumber
\end{equation}

\noindent and the result of this subtraction at NLO is to
eliminate the $\alpha_2$ term, and to transform the NLO
coefficients to the form of Table 5, just as in the case of the $K
\to\pi$ subtraction. This is exactly what is required, since the
NLO coefficients that appear in Table 5 always appear in the
transformed (finite) combinations in physical quantities, such as
$K\to\pi\pi$ at physical kinematics.

There is a subtlety in computing $K\to\pi\pi$ at the kinematics
where $m_K=m_\pi$ (UK1) that must be considered when the matrix
element of the $\Theta^{(3,\overline{3})}$ operator is computed.
One expects the power divergence in $\langle \pi^+\pi^-|{\cal
O}^{(8,1)}|K^{0}\rangle$ to vanish at $m_K=m_\pi$ by CPS
arguments, as discussed in \cite{berntwo}. However, a naive
calculation of $\langle
\pi^+\pi^-|\Theta^{(3,\overline{3})}|K^{0}\rangle$ in Minkowski
space shows that a factor of $m_K-m_\pi$ appears in the
denominator, potentially cancelling the $m_K-m_\pi$ in the
coefficient multiplying $\langle
\pi^+\pi^-|\Theta^{(3,\overline{3})}|K^{0}\rangle$ in Eq (67).
This point was clarified in \cite{goltsix}. As they point out, it
is crucial to use the $\epsilon$ prescription in order to have a
well defined Minkowski space amplitude.  The $\epsilon \rightarrow
0$ limit must be taken after the $m_K \rightarrow m_\pi$ limit.
Thus, at leading order in ChPT, the $K\to\pi\pi$ amplitude for the
$\Theta^{(3,\overline{3})}$ operator at UK1 is

\begin{equation}
\langle
\pi^+\pi^-|\Theta^{(3,\overline{3})}|K^{0}\rangle=\lim_{\epsilon\rightarrow
0}\left[\lim_{m_K\rightarrow m_\pi}
\frac{4i\alpha^{(3,\overline{3})}}{3f^3}\frac{3m_\pi(2m_\pi-m_K)+i\epsilon}{4m_\pi(m_K-m_\pi)-i\epsilon}\right].
\end{equation}

\noindent As one can see from Eq (68) after taking the first
limit, there is a pole at this kinematics. This pole in the
denominator comes about from the graph of Fig 1, C2, where the
kaon is annihilated by the tadpole operator.  As discussed by
\cite{goltsix}, it is useful to consider the corresponding
amplitude in finite volume Euclidean space.  In this case, the
divergence is regulated by the finite time extent of the lattice,
and the amplitude becomes proportional to $t_\pi$ (to LO in ChPT),
which is the difference in time between the weak operator
insertion and the two pion sink.  When one multiplies $\langle
\pi^+\pi^-|\Theta^{(3,\overline{3})}|K^{0}\rangle$ by
$m^2_K-m^2_\pi$ at exactly $m_K=m_\pi$, then the contribution from
the $\Theta^{(3,\overline{3})}$ term in Eq (67) is identically
zero, and the power divergence vanishes at $m_K=m_\pi$, as
expected from CPS symmetry.  In the previous version of this
paper, as well as in \cite{laiho}, we did not properly appreciate
this subtlety; we correct Eqs (31) and (D6) of \cite{laiho} in
Appendix F.

The subtraction is necessary in UKX at all accessible values of
the meson masses except UK1, as discussed above \footnote{Note
that UK1 is only accessible in the full theory \cite{linthree}.},
and at $m_K=2m_\pi$ (UK2) because then there is no 4-momentum
insertion at the weak vertex. To the extent that one cannot set
$m_K$ exactly equal to $2m_\pi$ on the lattice, it becomes
necessary to perform a small subtraction at this kinematics. Since
the power divergences in UKX are proportional to $m_K-2m_\pi$ [Eqs
(65, E10)], the best place to investigate UKX numerically is in
the vicinity of $m_K=2m_\pi$, where we hope the power divergences
will not be intractable.

The NLO LEC's for the (8,1) case can be obtained as follows.  From
the subtracted $K\to\pi$ amplitude, Eq (62), one can obtain the
leading order LEC, $\alpha_1$. If one uses the LEC combinations
obtainable from Eq (61) for $K \rightarrow 0$, $e^r_{2,rot}$ and
$e^r_{1,rot}-e^r_{5,rot}$, one can also obtain from Eq (62):
$e^r_{1,rot}+e^r_{3,rot}-4e^r_{39}$, $e^r_{10,rot}-e^r_{35}$,
$2e^r_{10,rot}-e^r_{11}$, and $e^r_{14}$. Using this information
along with the linear combinations one can get from Eq (66) (after
the subtraction) it is possible to obtain $e_{11}+2e_{15,rot}$ and
$e_{13,rot}$.  Notice that UKX provides additional redundancy over
that of UK1 and UK2 alone.  For the construction of the physical
$K \rightarrow \pi \pi$ amplitude we need these seven linear
combinations:
$[e_{2,rot},e_{1,rot}+e_{3,rot}-4e_{39},e_{10,rot}-e_{35},2e_{10,rot}-e_{11},e_{11}+2e_{15,rot},e_{13,rot},e_{14}]$.
One can verify that with these linear combinations it is possible
to obtain the linear combinations in Eq (58).

The logarithmic and Gasser-Leutwyler counterterm contributions to
the amplitudes presented in this section are given in Appendix E.

\section{Checks of the Calculations}

The logarithmic terms in the Appendixes of this paper are rather
lengthy, and so checks are important.  The first check these
expressions must pass is that the divergences from the one-loop
insertions cancel those of the divergent counterterms.  This was
checked for all expressions in this paper.  Another check is that
an expression reduces to some other in the appropriate limit. For
example, in the SU(3) limit, the equations in Appendix C reduce to
those of \cite{bijnens} in the same limit, as well as those of
\cite{golt}, modulo renormalization scheme dependent constants.
That is, the logarithmic terms agree, but the scheme dependent
$m^4$ coefficients differ.

The $K \to \pi\pi$ amplitudes in the full theory for the (8,8)'s
in Appendix D agree with Pallante, \etal \cite{pallante}, as well
as with \cite{cirig} (where only numerical values were given).
Also, the PQ, $K\to\pi$ amplitudes of Appendix D for the (8,8)'s
agree with \cite{cirig} when they reduce to those of the full
theory, in the SU(3) limit with $m_{sea}=m_{val}$.  In the
partially quenched theory, \cite{goltthree} has done $K^{+}\to
\pi^{+}$ in the SU(3) limit.  By taking the appropriate linear
combinations of the $\Delta I=3/2$ and 1/2 amplitudes given in
Appendix D we can compare to this special case, where we find
agreement.  In Appendix E, Eq (E1) [PQ $K\to 0$ for the (8,1)'s]
can be compared directly with \cite{golt}, where it agrees to
within renormalization scheme dependent constants. Eq (E2) [PQ
$K\to\pi$ for the (8,1)'s] also agrees with \cite{golt} in the
SU(3) limit modulo the renormalization scheme dependent constants.
Eq (E3), $K\to\pi\pi$ at UK1, reduces to that of the full theory
for $m_{SS}=m$, $N=3$, and can be compared with our previous paper
\cite{laiho}.  Note, however, there is an error in this quantity
in \cite{laiho} which has been corrected in Appendix F of this
paper.  Eq (E5), $K \to\pi\pi$ at UK2, does not reduce to that of
the full theory since $m_s\neq m_d=m_u$, but the sea quarks were
taken to be degenerate.  Note that the logarithmic parts of the
$\alpha_2$ term vanish for this on-shell quantity just as in the
full theory.

\section{Conclusion}

This paper demonstrates that all of the ingredients necessary to
construct all of the $K\to\pi\pi$ amplitudes to NLO in the full
theory can be obtained without 3-momentum insertion on the
lattice, which reduces the computational cost of obtaining the NLO
LEC's; all that is necessary in the needed $K\to \pi$ amplitudes
is the use of non-degenerate quark masses such that $m^{lat}_K
\neq m^{lat}_\pi$. It was also demonstrated that all of the
ingredients needed to produce $\epe$ to NLO are obtainable from
partially quenched ChPT.  In the case that $N=3$, the LEC's are
those of the full theory \cite{sharpe}.\footnote{By studying
finite volume Euclidean Green's functions in PQChPT,
\cite{linthree} confirmed that one can use the partially quenched
theory for $\Delta I =1/2$, $K\to\pi\pi$ amplitudes for $N=2$, and
our choice for the sea quark mass, $m_{sea}=m_{u,d}$.  However,
enhanced finite volume effects are present in the $N=3$ case
unless the sea quark masses are pairwise equal to the valence
quark masses, such that one is in the full theory.} The partially
quenched amplitudes were calculated under the assumption that both
the valence and sea quark masses are small compared to the $\eta'$
mass so that the $\eta'$ can be integrated out.  This means that
the $N=0$ limit of our amplitudes are not those of the quenched
approximation, and this has been discussed elsewhere \cite{golt}.
(The terms inversely proportional to powers of $N$ become quenched
chiral logs.) We point out that we are using the PQS method
\cite{goltfour}, where only the sea quarks propagate in the loops
of Fig 3c (See Section 6.1 for the correspondence between Fig 3c
and the traditional form of the eye-diagrams in Fig 2 for the
various operators), and that the necessary ingredients to obtain
the (8,1)'s are essentially unchanged from \cite{laiho} in this
prescription.\footnote{The only change is that instead of UK1, one
may need to use UKX, as the former has difficulties which were
pointed out by \cite{linthree}.  See also our note added in
revision.} The PQN method, however, may not be adequate to
determine the LEC's to NLO using only the ingredients needed of
the full theory, except in the case where it becomes the full
theory ($N=3$, $m_{sea}=m_{val}$).

The PQChPT formulas in this paper are valid for $N=2$, however,
and this should be useful for the work in progress by RBC with
$N=2$ dynamical flavors of domain wall quarks \cite{taku}, though
the values of the LEC's determined from these calculations will
not necessarily be those of the full theory.  One would hope, of
course, that the $N$ dependence will not be so severe, and that
this calculation will not be so far from the full theory.
Ultimately, one would like to check this with a full $N=3$
calculation.

We show how the bilinear $(3,\overline{3})$ operator is used to
eliminate the power divergences due to mixing with lower
dimensional operators to all orders in (PQ)ChPT for the $\Delta
I=1/2$ amplitudes.  The subtraction is performed to all orders in
ChPT \cite{blum}.  This is important, because the higher order
power divergent parts can overwhelm the physical terms one is
trying to calculate.

We have pointed out that the (8,8) $K\to\pi\pi$ amplitudes can be
constructed to NLO using only partially quenched $K\to\pi$
amplitudes with degenerate valence quark masses and without
momentum insertion.  Also, we showed how $K\to0$ can be used to
perform the $\Delta I=1/2$ power divergent subtraction. $K\to\pi$
calculations with nondegenerate valence quark masses would provide
additional redundancy in obtaining the needed NLO LEC's.

Finally, we point out that the threshold divergences that lead to
enhanced finite volume corrections to the lattice calculations of
$\Delta I=1/2$, $K\to\pi\pi$ amplitudes at NLO vanish in the
Minkowski space amplitudes considered in this paper when the sea
quark mass is equal to the up and down quark masses
($m_{sea}=m_u=m_d$).  This conclusion remains true in the $N=2$
case for the finite volume Euclidean correlation functions that
are relevant for lattice simulations, as demonstrated by
\cite{linthree}.  The $N=3$ case had been shown to be problematic
due to the presence of enhanced finite volume effects unless one
is working in the full theory, with all sea quark masses equal to
the corresponding valence quark masses \cite{linthree}.  See the
note added in revision for further discussion of this issue.

\section{Note added in revision}

Since the original posting of version 1 of this paper on the
preprint archive, Lin, \etal \cite{linthree} have submitted a
paper motivated at least in part by our version 1. They have done
a calculation of the relevant finite volume Euclidean correlation
functions, and they have made several important observations
regarding our attempts to obtain $K\to\pi\pi$ amplitudes at NLO,
which we briefly review, and then we make some comments.  They
point out that the case of the $\Delta I=1/2$ $K\to\pi\pi$
amplitudes at degenerate quark masses (what we call UK1), has
difficulties in the full theory, and is intractable in the
partially quenched theory, even at our special kinematics,
$m_{sea}=m_{u,d}$.

The difficulty with UK1 in the full theory is that one must
disentangle various two meson final states in order to obtain the
two pion final state.  These two meson states have different
energies, and since they appear in the correlation function with
different exponentials in time, they will be difficult to obtain.
See \cite{linthree} for further details.  They have also
discovered that when unphysical degrees of freedom propagating in
the meson re-scattering diagram are light enough to go on shell,
they cause enhanced finite volume effects, which cannot be
eliminated by going to larger lattices, and they make the
extraction of such amplitudes impossible in the infinite volume
limit. This problem afflicts the $\Delta I=1/2$ $K\to\pi\pi$
amplitudes for degenerate quark masses (UK1) in the partially
quenched theory, and so it cannot be used in our attempts to get
the LEC's to NLO.

However, it was also pointed out in \cite{linthree} that the
enhanced finite volume effects do vanish when the unphysical
degrees of freedom are heavier than the light quark mass.  This
leads to the conclusion that when one is working at UKX (initial
and final mesons at rest) if $m_K$ is strictly greater than
$m_\pi$, assuming also $m_{sea}=m_{u,d}$ and $N=2$, the enhanced
finite volume effects vanish.  Thus UK2 ($m_K=2m_\pi$), and the
kinematics points of UKX (with $m_K>m_\pi$) are useable in the
partially quenched theory.  Also, they point out that since the
unphysical degrees of freedom cannot be lighter than the light
quark mass, if one is working in the $N=3$ theory, one must use
full QCD.  That is, the idea of varying the sea and valence quark
masses independently \cite{sharpe} will not work for $\Delta
I=1/2$, $K\to\pi\pi$ amplitudes without introducing enhanced
finite volume effects.

As discussed in \cite{linthree}, it is not possible to use UK1 in
order to obtain some of the NLO LEC's in the partially quenched
case. We suggest that there may be a window where the quark masses
are light enough and the lattice size is small enough so that the
formulas of finite volume partially quenched ChPT can be used to
extract LEC's from numerical data.  Whether or not this proves
feasible, we have found that we do not need the information from
UK1 if we use the $K\to\pi\pi$ kinematics accessible to the
lattice that we call UKX. We have presented results at UKX (all
three mesons at rest, in general, requiring energy insertion with
$m_K \geq m_\pi$) in section 8, of which UK1 and UK2 are special
cases. According to \cite{linthree}, there will not be enhanced
finite volume effects at this kinematics (when $N=2$,
$m_{sea}=m_{u,d}$ and $m_K>m_\pi$), though, in general, one will
need to do the power divergent subtractions, as in the $K\to\pi$
case. We demonstrated in Section 8 that one can obtain all of the
LEC's needed for the (8,1), $K\to\pi\pi$ amplitudes using the UKX
kinematics points, along with the LEC's obtainable from $K\to0$
and $K\to\pi$.  We conclude that it is possible to obtain all of
the needed LEC's in the partially quenched theory for $N=2$,
though, as noted by \cite{linthree}, an $N=3$ determination will
require the full theory for the information needed from lattice
(8,1), $K\to\pi\pi$ amplitudes in order to avoid the enhanced
finite volume effects.

\bigskip
\bigskip
\bigskip
\bigskip

\centerline{\bf ACKNOWLEDGEMENTS}
\bigskip

This research was supported in part by US DOE Contract No.
DE-AC02-98CH10886.  We thank Tom Blum, Chris Dawson, Maarten
Golterman, David Lin and Kim Splittorff for discussions.

\bigskip
\bigskip
\bigskip

\appendix
\section*{\bf APPENDIX A}
\setcounter{equation}{0} \setcounter{section}{1}
\renewcommand{\theequation}{A\arabic{equation}}

\bigskip

Appendixes B-E contain the finite logarithm and Gasser-Leutwyler
counterterm contributions to the amplitudes presented in this
paper. They were calculated using the \textsc{FeynCalc} package
\cite{mert} written for the \textsc{Mathematica} \cite{wolfram}
system.  These expressions involve the regularized
Veltman-Passarino basis integrals $A_0$, $B_0$ and $C_0$
\cite{pass}:

\be A_0(m^2) = \frac{1}{16\pi^2f^2} m^2 \ln \frac{m^2}{\mu^2}, \ee

\bea B_0(q^2,m^2_1,m^2_2)\!\!\!\!\! & = &\!\!\!\!\! \int^1_0{dx}
\frac{1}{(4\pi f)^2} [1+\ln (-x(1-x) q^2 +xm^2_1
+(1-x)m^2_2)\nonumber \\
\!\!\!\!\!& &\!\!\!\!\! -\ln \mu^2 ], \eea

\bea C_0(0,q^2,q^2,m^2_1,m^2_1,m^2_2)\!\!\!\!\! & = &\!\!\!\!\!
\frac{1}{(4\pi f)^2} \int^1_0{dx}\frac{x}{-x(1-x) q^2 +xm^2_1
+(1-x)m^2_2}. \nonumber \\ \eea

Note that the original Veltman-Passarino integrals did not involve
ChPT, and so the pseudoscalar decay constant, $f$, is not part of
the original definitions of the integrals, but is inserted here
for convenience.

\appendix
\section*{\bf APPENDIX B}
\setcounter{equation}{0} \setcounter{section}{1}
\renewcommand{\theequation}{B\arabic{equation}}
\bigskip

At 1-loop order in the partially quenched theory the pseudoscalar
decay constants and masses are renormalized such that
$f_{\pi,K}=f\left(1+\frac{\Delta f_{\pi,K}}{f}\right)$ and
$m^2_{\pi,K (1-loop)}=m^2_{\pi,K}\left(1+\frac{\Delta
m^2_{\pi,K}}{m^2_{\pi,K}}\right)$.  The corrections are

\bea \frac{\Delta f_\pi}{f} & = & -NA_0(m^2_{uS}) + \frac{8}{f^2}
(L_5 m^2_{\pi}+L_4 Nm^2_{SS}), \\
\frac{\Delta f_K}{f} & = &
\frac{1}{N16\pi^2f^2}(m^2_K-m^2_{SS})-\frac{m^4_{\pi}+m^2_K(m^2_{SS}-2m^2_{\pi})}{2N(m^2_K-m^2_{\pi})}
\left(\frac{1}{m^2_{\pi}}A_0(m^2_\pi) -
\frac{1}{m^2_{33}}A_0(m^2_{33})\right) \nonumber \\ &&
-\frac{N}{2}( A_0(m^2_{uS})+A_0(m^2_{sS})) +\frac{8}{f^2} (L_5
m^2_K + L_4 Nm^2_{SS}), \nonumber \\ \eea

\bea \frac{\Delta m^2_\pi}{m^2_\pi} & = &
\frac{2}{N}\left[\frac{1}{16 \pi^2 f^2}(-m^2_{SS}+m^2_{\pi})
+\frac{2m^2_{\pi}-m^2_{SS}}{m^2_{\pi}}A_0(m^2_\pi)\right]  -
\frac{16}{f^2} [(L_5-2L_8) m^2_\pi  \nonumber \\
& & + (L_4-2L_6) Nm^2_{SS} ],  \\
\frac{\Delta m^2_K}{m^2_K}  & = &
\frac{-1}{N(m^2_K-m^2_{\pi})}\left[(m^2_{\pi}-m^2_{SS})
A_0(m^2_\pi)+(-2m^2_K+m^2_\pi+m^2_{SS}) A_0(m^2_{33})\right]
\nonumber \\ &&
 -\frac{16}{f^2} [(L_5-2L_8) m^2_K +
(L_4-2L_6) Nm^2_{SS}]. \eea

\noindent For degenerate quark masses at 1-loop order, $m^{2}_{K
(1-loop)}=m^{2}_{\pi (1-loop)}=m^{2}\left(1+\frac{\Delta
m^2}{m^2}\right)$, $f_\pi=f_K=f\left(1+\frac{\Delta f}{f}\right)$,

\bea \frac{\Delta m^2}{m^2} & = &
\frac{2}{N}\left[\frac{m^2-m^2_{SS}}{16\pi^2f^2}+\frac{2m^2-m^2_{SS}}{m^2}
A_0(m^2)\right] - \frac{16}{f^2} [(L_5 - 2L_8)m^2 \nonumber \\ &&
+ (L_4-2L_6) Nm^2_{SS}], \\
\frac{\Delta f}{f} & = & -NA_0(m^2_{vS}) + \frac{8}{f^2} (L_5 m^2
+ L_4 Nm^2_{SS}), \eea

\bigskip

with $m^2_{vS} = \frac{1}{2}(m^2+m^2_{SS})$.

\bigskip
\bigskip

\bigskip
\bigskip
\bigskip

\appendix
\section*{\bf APPENDIX C: Log Corrections to full ChPT}
\setcounter{equation}{0} \setcounter{section}{1}
\renewcommand{\theequation}{C\arabic{equation}}
\bigskip

The logarithmic corrections to the $K \rightarrow \pi$ amplitudes
in the full theory when 3-momentum insertion vanishes are

\bea \bra \pi^+ |{\cal O}^{(27,1),(3/2)} |K^+\ket_{log} & = &
-\frac{4\alpha_{27}}{f^2} m_K m_\pi \left[ -2m_K m_\pi B_0(q^2,
m^2_K, m^2_\pi) - \frac32 A_0(m^2_\eta) - 7A_0(m^2_K) \right. \nonumber \\
& & \left.- \frac{15}{2} A_0(m^2_\pi)- \frac{\Delta f_K}{f} -
\frac{\Delta f_\pi}{f} + \frac{1}{2}\left(\frac{\Delta
m^2_K}{m^2_K}+\frac{\Delta m^2_{\pi}}{m^2_{\pi}}\right) \right],
\eea

\bea \bra \pi^+|{\cal O}^{(27,1),(1/2)} |K^+\ket_{\log} & = &
-\frac{4\alpha_{27}}{f^2}m_K m_\pi \biggl[-(4m^2_K+6 m_K m_\pi
-4m^2_\pi) B_0(q^2,m^2_K, m^2_\eta) \biggr. \nonumber \\ && + 4m_K
m_\pi B_0(q^2, m^2_K, m^2_\pi) + \frac{3(4m^2_K -3m_K
m_\pi+2m^2_\pi)}{2m_\pi(m_K-m_\pi)} A_0(m^2_\eta)\nonumber \\
& & -  \frac{6m^2_K +10m_Km_\pi -10m^2_\pi}{m_\pi(m_K-m_\pi)}
A_0(m^2_K) - \frac{3(m_K-2m_\pi)}{2(m_K-m_\pi)} A_0(m^2_\pi) \nonumber \\
&&\left. -  \frac{\Delta f_K}{f} - \frac{\Delta f_\pi}{f} +
\frac{1}{2}\left(\frac{\Delta m^2_K}{m^2_K}+\frac{\Delta
m^2_{\pi}}{m^2_{\pi}}\right)
 \right], \eea

 \bea \bra\pi^+ | {\cal O}^{(8,1)} |K^+ \ket_{log} & = &
\frac{4\alpha_1}{f^2}m_K m_\pi \left[\frac{1}{9}(4m^2_K+6m_K
m_\pi-4m^2_\pi ) B_0(q^2,m^2_K,m^2_\eta) \right.
\nonumber \\
& & + 4m_K m_\pi B_0(q^2,m^2_K, m^2_\pi) - \frac{4m^2_K+7m_K m_\pi
- 8m^2_\pi}{6m_\pi(m_K-m_\pi)}
A_0(m^2_\eta) \nonumber \\
& & - \frac{3m^2_K+5m_Km_\pi -5m^2_\pi}{3m_\pi(m_K-m_\pi)}
 A_0(m^2_K)   -\frac{3(m_K-2m_\pi)}{2(m_K-m_\pi)} A_0(m^2_\pi) \nonumber \\ &&
\left. -\frac{\Delta f_K}{f} - \frac{\Delta f_\pi}{f} +
\frac{1}{2}\left(\frac{\Delta m^2_K}{m^2_K}+\frac{\Delta
m^2_{\pi}}{m^2_{\pi}}\right)\right] \nonumber \\
& & -\frac{4\alpha_2}{f^2} m^2_K \left[ \frac{2}{3}m_K m_\pi
B_0(q^2,
m^2_K, m^2_\eta) \right. + 4 m_K m_\pi B_0(q^2, m^2_K, m^2_\pi)\nonumber \\
& & -\frac{5m_K-2m_\pi}{6(m_K-m_\pi)} A_0(m^2_\eta) - \frac{3m_K -
2m_\pi}{m_K-m_\pi} A_0(m^2_K) \nonumber
\\
& & \left. - \frac{3m_K-6m_\pi}{2(m_K-m_\pi)} A_0(m^2_\pi) -
\frac{\Delta f_k}{f} - \frac{\Delta f_\pi}{f} \right]. \eea

These are the simplified versions of \cite{laiho}, Eqs (C2), (D3)
and (D4), respectively, when $q^2=(m_K-m_\pi)^2$.

\bigskip
\bigskip
\bigskip

\appendix
\section*{\bf APPENDIX D: PQ Log Corrections to (8,8)'s}
\setcounter{equation}{0} \setcounter{section}{1}
\renewcommand{\theequation}{D\arabic{equation}}
\bigskip

The logarithmic corrections for the (8,8) amplitudes relevant for
the determination of $K \rightarrow \pi \pi$ are given in this
section.  The logarithmic corrections to $K \rightarrow \pi \pi$
in the full theory were calculated first in \cite{cirig,pallante},
and are included here for completeness.

\bea \bra\pi^+\pi^- | {\cal O}^{(8,8),(3/2)} | K^0\ket_{log} & = &
-4i \frac{\alpha_{88}}{f_Kf^2_\pi} \left[ \left(
\frac{5m^4_K}{4m^2_\pi} -2m^2_K\right) B_0(m^2_\pi, m^2_K, m^2_\pi)\right. \nonumber \\
& & +(m^2_K-2m^2_\pi) B_0(m^2_K, m^2_\pi, m^2_\pi) \nonumber \\
& & + \frac{m^4_K}{4m^2_\pi} B_0(m^2_\pi, m^2_K,
m^2_\eta)\nonumber
- \left(4+ \frac{m^2_K}{2m^2_\pi}\right) A_0(m^2_K)\nonumber \\
& & + \left( \frac{5m^2_K}{4m^2_\pi} -8 \right )
A_0(m^2_\pi)\left. - \frac{3m^2_K}{4m^2_\pi} A_0(m^2_\eta)\right],
\eea

\bea \bra\pi^+\pi^- | {\cal O}^{(8,8),(1/2)} | K^0\ket_{log} & = &
-8i \frac{\alpha_{88}}{f_Kf^2_\pi} \left[ \left(
\frac{m^4_K}{2m^2_\pi} -2m^2_K\right) B_0(m^2_\pi, m^2_K, m^2_\pi)\right. \nonumber \\
& &+\frac{3}{4}m^2_K B_0(m^2_K,m^2_K,m^2_K) +(m^2_\pi-2m^2_K) B_0(m^2_K, m^2_\pi, m^2_\pi) \nonumber \\
& & + \frac{m^4_K}{4m^2_\pi} B_0(m^2_\pi, m^2_K,
m^2_\eta)\nonumber
+\frac{1}{4} \left(\frac{m^2_K}{m^2_\pi}-22 \right) A_0(m^2_K)\nonumber \\
& & + \frac{1}{4}\left( \frac{2m^2_K}{m^2_\pi} -26 \right )
A_0(m^2_\pi)\left. - \frac{3m^2_K}{4m^2_\pi} A_0(m^2_\eta)\right].
\eea

The $\Delta I=3/2$, $K \rightarrow \pi$ logarithmic corrections
are given by

\bea \bra \pi^+|{\cal O}^{(8,8),(3/2)} |K^+\ket_{\log} & = &
\frac{4\alpha_{88}}{f^2} \biggl[-2m_K m_\pi B_0(q^2,m^2_K,
m^2_\pi)
\biggr.  -N(A_0(m^2_{sS})+3A_0(m^2_{uS}))\nonumber \\
& & + \frac{m^4_\pi
+m^2_K(m^2_{SS}-2m^2_\pi)}{N(2m^4_K-3m^2_Km^2_\pi+m^4_\pi)}
A_0(m^2_{33}) \nonumber \\ && +
\frac{m^4_\pi+m^2_K(m^2_{SS}-2m^2_\pi)}{Nm^2_\pi(m^2_\pi-m^2_K)}
A_0(m^2_\pi) +\frac{2}{N16 \pi^2 f^2}(m^2_K-m^2_{SS}) \nonumber \\
&&\left.-  \frac{\Delta f_K}{f} - \frac{\Delta f_\pi}{f}
 \right]. \eea

For the $\Delta I =1/2$, $K \rightarrow \pi$ corrections there are
(at least) two possibilities, when the electroweak operator is
partially quenched and when it is not.  The following amplitude
corresponds to quenching the short distance electroweak operator,
neglecting the type of contraction in Fig 3c (with a photon or Z
replacing the gluon),

\bea \bra \pi^+|{\cal O}^{(8,8),(1/2)} |K^+\ket_{\log} & = &
\frac{8\alpha_{88}}{f^2} \biggl[m_K m_\pi B_0(q^2,m^2_K, m^2_\pi)
\biggr. + Nm_K m_\pi B_0(q^2, m^2_{uS}, m^2_{sS})\nonumber \\ && +
\frac{N(2m_\pi -3m_K)}{2(m_K-m_\pi)} A_0(m^2_{sS})+ \frac{N(6m_\pi
-5m_K)}{2(m_K-m_\pi)} A_0(m^2_{uS})\nonumber \\
& & + \frac{m^4_\pi
+m^2_K(m^2_{SS}-2m^2_\pi)}{N(2m^4_K-3m^2_Km^2_\pi+m^4_\pi)}
A_0(m^2_{33}) \nonumber \\ && +
\frac{m^4_\pi+m^2_K(m^2_{SS}-2m^2_\pi)}{Nm^2_\pi(m^2_\pi-m^2_K)}
A_0(m^2_\pi) +\frac{2}{N16 \pi^2 f^2}(m^2_K-m^2_{SS}) \nonumber \\
&&\left.-  \frac{\Delta f_K}{f} - \frac{\Delta f_\pi}{f}
 \right], \eea

\noindent while the next corresponds to where the short distance
electroweak operator is not quenched, and valence quarks do
propagate in the loops of Fig 3c (again with a photon or Z
replacing the gluon).

\bea \bra \pi^+|{\cal O}^{(8,8),(1/2)} |K^+\ket_{\log} & = &
\frac{8\alpha_{88}}{f^2} \biggl[2m_K m_\pi B_0(q^2,m^2_K, m^2_\pi)
-m_K m_\pi B_0(q^2,m^2_K, m^2_{33}) \biggr. \nonumber \\ && + Nm_K
m_\pi B_0(q^2, m^2_{uS}, m^2_{sS}) +
\frac{N(2m_\pi -3m_K)}{2(m_K-m_\pi)} A_0(m^2_{sS})\nonumber \\
& & + \frac{N(6m_\pi -5m_K)}{2(m_K-m_\pi)} A_0(m^2_{uS})+\biggl(
\frac{m^4_\pi
+m^2_K(m^2_{SS}-2m^2_\pi)}{N(2m^4_K-3m^2_Km^2_\pi+m^4_\pi)}
\nonumber \\ && \biggr. \biggl. +\frac{m_K}{2(m_K-m_\pi)}\biggr)
A_0(m^2_{33})
+\biggl(\frac{m^4_\pi+m^2_K(m^2_{SS}-2m^2_\pi)}{Nm^2_\pi(m^2_\pi-m^2_K)}
\nonumber \\ && \biggr. \biggl.
+\frac{m_K}{2(m_K-m_\pi)}\biggr)A_0(m^2_\pi)+\frac{m_K}{m_\pi-m_K}
A_0(m^2_K)
+\frac{2}{N16 \pi^2 f^2}(m^2_K-m^2_{SS}) \nonumber \\
&&\left.-  \frac{\Delta f_K}{f} - \frac{\Delta f_\pi}{f}
 \right]. \eea

 In the case of degenerate valence quarks the above expressions
 reduce to

\bea \bra \pi^+|{\cal O}^{(8,8),(3/2)} |K^+\ket_{\log} & = &
\frac{4\alpha_{88}}{f^2} \left\{\frac{-2}{16\pi^2 f^2}\biggl[ m^2
\ln\frac{m^2}{\mu^2}+N(m^2+m^2_{SS})
\ln\biggl(\frac{m^2+m^2_{SS}}{2\mu^2} \biggr) \biggr. \right. \nonumber \\
&& \left. \biggl. +m^2 \biggr] - \frac{2\Delta f}{f} \right\},
\eea

\bea \bra \pi^+|{\cal O}^{(8,8),(1/2)} |K^+\ket_{\log} & = &
\frac{8\alpha_{88}}{f^2} \left\{\frac{1}{16\pi^2 f^2}\biggl[ m^2
\ln\frac{m^2}{\mu^2}-2N(m^2+m^2_{SS})
\ln\biggl(\frac{m^2+m^2_{SS}}{2\mu^2} \biggr) \biggr. \right. \nonumber \\
&& \left. \biggl. +m^2 \biggr] - \frac{2\Delta f}{f} \right\}.
\eea

\noindent Note that the two $\Delta I=1/2$ expressions reduce to
the same thing in the SU(3) limit.

\bigskip
\bigskip
\bigskip

\appendix
\section*{\bf APPENDIX E:  PQ Log Corrections to (8,1)'s}
\setcounter{equation}{0} \setcounter{section}{1}
\renewcommand{\theequation}{E\arabic{equation}}
\bigskip

The logarithmic corrections for the quantities relevant for the
determination of the (8,1), $K \rightarrow \pi \pi$ amplitudes are
given in this Appendix.  The logarithmic corrections to the
physical $K \rightarrow \pi \pi$ amplitude have been done by
\cite{kambor,ecker}, and we refer to \cite{laiho}, Eq. D10, for
the amplitude in our conventions.

The logarithmic corrections to $K \rightarrow 0$ and $K
\rightarrow \pi$ are given by

\bea \bra 0|{\cal O}^{(8,1)}|K^0\ket_{log}\!\!\!\! & = &\!\!\!\!
\frac{4i\alpha_2}{f} (m^2_K-m^2_\pi) \left[
-N(A_0(m^2_{sS})+A_0(m^2_{uS}))
+\frac{-4m^2_K+2m^2_\pi+m^2_{SS}}{N(m^2_{\pi}-2m^2_K)}A_0(m^2_{33})\right.\nonumber
\\ &&  \left. +\frac{2m^2_\pi-m^2_{SS}}{Nm^2_\pi}
A_0(m^2_\pi)-\frac{2(m^2_{SS}-m^2_K)}{N16 \pi^2 f^2} -\frac{\Delta f_K}{f}\right] \nonumber \\
&& + \frac{4i\alpha_1}{f}
\left[\frac{-2(m^2_K-m^2_\pi)(2m^2_K-m^2_{SS})}{N16\pi^2f^2}
\right. \nonumber \\ && +
N(m^2_{sS}A_0(m^2_{sS})-m^2_{uS}A_0(m^2_{uS}))
+\frac{1}{N}(3m^2_\pi-2m^2_{SS}) A_0(m^2_\pi)\nonumber \\ &&
\left. + \frac{1}{N} (2m^2_{SS}-3m^2_{33}) A_0(m^2_{33}) \right],\nonumber \\
&& \eea

 \bea \bra\pi^+ | {\cal O}^{(8,1)} |K^+ \ket_{log} & = &
\frac{4\alpha_1}{f^2}m_K m_\pi \left[\frac{4m^3_K+4m^2_Km_\pi-2m_K
m^2_{SS}+m^3_\pi-3m_\pi m^2_{SS}}{N16 \pi^2 f^2 m_\pi} \right.
\nonumber \\ && -\frac{2}{N}(m^2_K+m_K
m_\pi-m^2_\pi)(2m^2_K-m^2_\pi-m^2_{SS})
C_0(0,q^2,q^2,m^2_{33},m^2_{33},m^2_K)\nonumber \\ &&
+\frac{1}{N(m^2_K-m^2_\pi)}\biggl(
 -6m^4_K-4m^3_K m_\pi+10m^2_K m^2_\pi+3m_K
m^3_\pi-4m^4_\pi \biggr. \nonumber \\ && \biggl. +(2m^2_K+m_K
m_\pi-2m^2_\pi)m^2_{SS}\biggr) B_0(q^2,m^2_K,m^2_{33})
\nonumber \\
& & + \frac{m_K m_\pi(m^2_\pi-m^2_{SS})}{N(m^2_K-m^2_\pi)}
B_0(q^2,m^2_K, m^2_\pi) \nonumber \\ && +N(-m^2_\pi+2m_K m_\pi +m^2_{SS}) B_0(q^2,m^2_{sS},m^2_{uS})\nonumber \\
&&- \frac{N(2m^2_K-m^2_\pi +m^2_{SS})}{2m_\pi(m_K-m_\pi)}
A_0(m^2_{sS})+ \frac{N(3m^2_\pi-2m_K m_\pi
+m^2_{SS})}{2m_\pi(m_K-m_\pi)}
A_0(m^2_{uS})\nonumber \\
& & - \frac{-12m^3_K-6m^2_K m_\pi+6m_K m^2_\pi+4m_K
m^2_{SS}+3m^3_\pi+ m_\pi m^2_{SS}}{2Nm_\pi(m^2_K-m^2_\pi)}
 A_0(m^2_{33})\nonumber \\ &&
 +  \frac{1}{2Nm^2_\pi(m^2_K-m^2_\pi)}\biggl(-7m^4_\pi-6m_K m^3_\pi+4m^2_K m^2_\pi
 +3m^2_\pi m^2_{SS} \biggr.\nonumber \\ && \biggl. +4m_K m_\pi m^2_{SS}-2m^2_K m^2_{SS}\biggr)
 A_0(m^2_{\pi})-\frac{2}{N} A_0(m^2_K) \nonumber \\ &&
 \left. -\frac{\Delta f_K}{f} - \frac{\Delta
f_\pi}{f} + \frac{1}{2}\left(\frac{\Delta
m^2_K}{m^2_K}+\frac{\Delta
m^2_{\pi}}{m^2_{\pi}}\right)\right] \nonumber \\
& & -\frac{4\alpha_2}{f^2} m^2_K \left[\frac{2}{N16 \pi^2
f^2}(m^2_K+m_K m_\pi+m^2_\pi-m^2_{SS}) \right. \nonumber \\
&& +\frac{2}{N}m_K
m_\pi(-2m^2_K+m^2_\pi+m^2_{SS})C_0(0,q^2,q^2,m^2_{33},m^2_{33},m^2_K)\nonumber
\\ && + \frac{m_K m_\pi(-4m^2_K+3m^2_\pi+m^2_{SS})}{N(m^2_K-m^2_\pi)} B_0(q^2,
m^2_K, m^2_{33})\nonumber \\ && + \frac{m_K
m_\pi(m^2_\pi-m^2_{SS})}{N(m^2_K-m^2_\pi)} B_0(q^2,
m^2_K, m^2_\pi)+ 2N m_K m_\pi B_0(q^2, m^2_{sS}, m^2_{uS})\nonumber \\
& &+\frac{N
m_K}{m_\pi-m_K}A_0(m^2_{sS})+\frac{N(m_K-2m_\pi)}{m_\pi-m_K}A_0(m^2_{uS})\nonumber
\\ && +\frac{m_K(4m^2_K-2m^2_\pi-m^2_{SS})}{N(m_K-m_\pi)(2m^2_K-m^2_\pi)}
A_0(m^2_{33})\nonumber \\ && +
\frac{(2m_\pi-m_K)(2m^2_\pi-m^2_{SS})}{Nm^2_\pi (m_\pi-m_K)}
A_0(m^2_\pi) \left.  - \frac{\Delta f_K}{f} - \frac{\Delta
f_\pi}{f} \right]. \eea
\medskip

\noindent In the case of degenerate valence quarks, the above
expression becomes

\bea \bra \pi^+|{\cal O}^{(8,1)} |K^+\ket_{\log} & = &
\frac{4\alpha_{1}}{f^2}m^2 \left\{\frac{1}{16\pi^2
f^2}\biggl[\frac{4}{N} (3m^2-m^2_{SS})
\ln\frac{m^2}{\mu^2}-\frac{3N}{2}(m^2+m^2_{SS}) \biggr. \right. \nonumber \\
&& \left. \biggl. \times \ln\biggl(\frac{m^2+m^2_{SS}}{2\mu^2}
\biggr) +\frac{2}{N}(5m^2-3m^2_{SS}) \biggr] - \frac{2\Delta
f}{f}+\frac{\Delta m^2}{m^2} \right\}\nonumber
\\ && -\frac{4\alpha_{2}}{f^2}m^2 \left\{\frac{1}{16\pi^2
f^2}\biggl[\frac{2}{N} (4m^2-m^2_{SS})
\ln\frac{m^2}{\mu^2}-N(m^2+m^2_{SS}) \biggr. \right. \nonumber \\
&& \left. \biggl. \times \ln\biggl(\frac{m^2+m^2_{SS}}{2\mu^2}
\biggr) +\frac{1}{N}(8m^2-4m^2_{SS}) \biggr] - \frac{2\Delta f}{f}
\right\}. \eea

The $K \rightarrow \pi \pi$ amplitude at UK1 in infinite volume
Minkowski space is given by

\bea \bra\pi^+\pi^- | {\cal O}^{(8,1)} | K^0\ket_{log} & = & 8i
\frac{\alpha_1}{f^3} m^2 \left\{ \frac{1}{16 \pi^2 f^2}
\biggl[\frac{2}{N}m^2_{SS}\ln \frac{m^2}{\mu^2} -3N m^2 \ln
\left(\frac{m^2+m^2_{SS}}{2 \mu^2}\right)\biggr. \right. \nonumber \\
&& -\frac{N}{2}(5m^2-m^2_{SS})\lambda_0 \nonumber \\ && + \lim_{s
\to 4m^2}\left(
 \frac{\pi i (m^2-m^2_{SS})((32N-3)m^2+3m^2_{SS})}{16N^2
m(s-4m^2)^{\frac12}} \right)+\frac{1}{8N}\nonumber \\ &&
 \biggr. \times[(22N^2-57)m^2+
m^2_{SS}(-2N^2+25)]\biggr]  \left. - \frac{3\Delta f}{f} +
\frac{\Delta m^2}{m^2} \right\}, \nonumber \\ &&  \eea

\noindent for $m_K=m_\pi=m$, where

\bea  \lambda_0=\frac{i}{\sqrt{2}}\sqrt{\frac{m^2_{SS}}{m^2}-1}
\ln
\left(\frac{\sqrt{\frac{m^2_{SS}}{m^2}-1}-\sqrt{2}i}{\sqrt{\frac{m^2_{SS}}{m^2}-1}+\sqrt{2}i}
\right) \eea

Expression E5 is real for $m^2_{SS}\geq m^2$.  When $m^2_{SS}<
m^2$, E5 has an imaginary part.  The $K \rightarrow \pi \pi$
amplitude at UK2 in infinite volume Minkowski space is

\bea \bra\pi^+\pi^- | {\cal O}^{(8,1)} | K^0\ket_{log} & = & 3i
\frac{\alpha_1}{f^3} m^2_K \left\{ \frac{1}{16 \pi^2 f^2}
\biggl[\frac{-m^2_K(5N^2-6N+8)}{2N^2}\ln \frac{m^2_K}{\mu^2}
\biggr. \right. \nonumber \\ && +\frac{N}{48}
\left(\frac{-16m^4_{SS}}{m^2_K}-8m^2_{SS}+3m^2_K \right) \ln
\left(\frac{7m^2_K+4m^2_{SS}}{\mu^2}\right)\nonumber \\ &&
+\frac{N}{48} \left(\frac{16m^4_{SS}}{m^2_K}-16m^2_{SS}-69m^2_K
\right) \ln
\left(\frac{m^2_K+4m^2_{SS}}{\mu^2}\right) \nonumber \\
&& -\frac{N}{12}(17m^2_K+4m^2_{SS})\lambda_1
-\frac{N}{6}(5m^2_K+4m^2_{SS})\lambda_2 \nonumber \\ && + \lim_{s
\to m^2_K}\left(\frac{-i \pi
m_K(m^2_K-4m^2_{SS})^2}{16N^2(s-m^2_K)^{\frac{3}{2}}} \right. \nonumber \\
&& \left.  +\frac{i \pi
(m^2_K-4m^2_{SS})(m^2_K(24N-43)-20m^2_{SS})}{96N^2
m_K(s-m^2_K)^{\frac12}} \right)\nonumber \\ &&
-\frac{(7m^2_K-4m^2_{SS})^2}{3\sqrt{6}N^2m^2_K}\cot^{-1}\left(\sqrt{6}\right)
-\frac{4\sqrt{3}m^2_K}{N}\tan^{-1}\left(\frac{2}{\sqrt{3}}\right)\nonumber
\\ &&
+\frac{1}{12\sqrt{3}N^2m^2_K}(7m^2_K-4m^2_{SS})[m^2_K(3N-10)+16m^2_{SS}]
\tan^{-1}\left(\frac{5}{\sqrt{3}}\right) \nonumber \\ &&
+\frac{\pi}{72\sqrt{3}N^2m^2_K}[(123N+70)m^4_K+4m^2_K
m^2_{SS}(3N-38)+64m^4_{SS}] \nonumber \\ &&
+\frac{\ln7}{72N^2m^2_K}[-(659N+252)m^4_K+4m^2_K
m^2_{SS}(107N-6)+96m^4_{SS}] \nonumber \\ &&
+\frac{\ln2}{24N^2}[(99N^3-40N^2+432N+192)m^2_K+12Nm^2_{SS}(3N^2-32)]
\nonumber \\ && +\frac{1}{432N^2}
[(-108N^3+1296N^2+2851N+2160)m^2_K
\nonumber \\ && \biggl. -4m^2_{SS}(108N^3+43N+432)]\biggr] \nonumber \\
&& \left. - \frac{\Delta f_K}{f} - \frac{2\Delta f_\pi}{f}
+\frac{4}{3m^2_K}\biggl(\Delta m^2_K-\Delta m^2_\pi\biggr) \right\}, \nonumber \\
& &   \eea

for $m_{K (1-loop)}=2m_{\pi (1-loop)}$, where

\bea  \lambda_1=i\sqrt{2\frac{m^2_{SS}}{m^2_K}-\frac12} \ln
\left(\frac{\sqrt{8\frac{m^2_{SS}}{m^2_K}-2}-2i}{\sqrt{8\frac{m^2_{SS}}{m^2_K}-2}+2i}
\right), \eea

\bea  \lambda_2=i\sqrt{2\frac{m^2_{SS}}{m^2_K}-\frac12} \ln
\left(\frac{\sqrt{8\frac{m^2_{SS}}{m^2_K}-2}+4i}{\sqrt{8\frac{m^2_{SS}}{m^2_K}-2}-4i}
\right). \eea

Note the imaginary threshold divergences in both (E4) and (E6). On
the lattice they are expected to contribute in the form of
enhanced finite volume effects.  See, for example, \cite{lintwo}.
When $m_{SS}< m_K/2=m_\pi$, Eqs. (E7) and (E8) have imaginary
parts.

Eq (E6) is most useful in fits to lattice data at the special
kinematics $m_{sea}=m_{u,d}$ ($m_{SS}=m_\pi$), $N=2$.  In this
case it reduces to

\bea \bra\pi^+\pi^- | {\cal O}^{(8,1)} | K^0\ket_{log} & = & 3i
\frac{\alpha_1}{f^3} m^4_K \left\{ \frac{1}{16 \pi^2 f^2}
\biggl[-5\ln \frac{m^2_K}{\mu^2}
-\sqrt{\frac{3}{2}}\cot^{-1}\left(\sqrt{6}\right) \biggr. \right.
\nonumber \\ && -2\sqrt{3}\tan^{-1}\left(\frac{2}{\sqrt{3}}
\right) +\frac{\pi}{\sqrt{3}} -\frac{113}{24}\ln7
+\frac{40}{3}\ln2 \nonumber \\ && +\frac{25}{4}
 \biggl. \biggr] \left. - \frac{\Delta f_K}{f} - \frac{2\Delta f_\pi}{f}
+\frac{4}{3m^2_K}\biggl(\Delta m^2_K-\Delta m^2_\pi\biggr)
\right\}. \nonumber \\ &&  \eea

The logarithmic contribution to UKX (kaon, pions at rest) in the
special case of $m_{sea}=m_u=m_d$ ($m_{SS}=m_\pi$) and $m_K>m_\pi$
is given below. In this expression, as in all others in this set
of Appendixes, $q^2=(m_K-m_\pi)^2$.

 \bea \bra\pi^+ \pi^- | {\cal O}^{(8,1)} |K^0 \ket_{log} & = &
\frac{4i\alpha_1}{f^3}m_\pi(m_K + m_\pi)
\left[\frac{m_K(2m^2_K+m^2_\pi)}{N m_\pi16 \pi^2 f^2} \right. \nonumber \\
&&
-\frac{4}{N}m^2_K(m^2_K-m^2_\pi)C_0(0,q^2,q^2,m^2_{33},m^2_{33},m^2_K)\nonumber
\\ && +\frac{1}{N}(-6m^2_K+4m_K m_\pi) B_0(q^2,m^2_K,m^2_{33}) +
2(N-2)m_K m_\pi B_0(q^2,m^2_K, m^2_\pi) \nonumber \\ &&
-4m_\pi(m_K-2m_\pi)
B_0(4m^2_\pi,m^2_K,m^2_K)+\frac{2}{N^2}(N-2)m_K(m_K-m_\pi)\nonumber
\\ && \times B_0(4m^2_\pi,m^2_\pi,m^2_{33}) -\frac{4}{N^2}m_\pi(2m_K-3m_\pi)
B_0(4m^2_\pi,m^2_{33},m^2_{33})  \nonumber \\
& & -\frac{2}{N^2}(N^3+2N^2-2N+2)m^2_\pi
B_0(4m^2_\pi,m^2_\pi,m^2_\pi)  \nonumber \\ && -
\frac{N^2m^3_K+4m^2_K m_\pi+2(N^2+2N-4)m_K m^2_\pi-12N
m^3_\pi}{2Nm_\pi(m^2_K-m^2_\pi)}
 A_0(m^2_K)\nonumber \\ &&
 -  \frac{1}{2N^2m_\pi(m^2_K-m^2_\pi)(2m^2_K-m^2_\pi)}\biggl(-12Nm^5_K +
  8(2N-1)m^4_K m_\pi \biggr. \nonumber \\ && \biggl. +4(N+6)m^3_K m^2_\pi -4(2N+3)m^2_K m^3_\pi +3(N-4)m_K m^4_\pi
  +8m^5_\pi \biggr) \nonumber \\ && \times A_0(m^2_{33})
  +\frac{1}{2N^2(m^2_K-m^2_\pi)} \biggl(-2(N^3-2N+2)m^2_K \nonumber \\ &&
  +(3N^3+4N^2-13N+12)m_K
  m_\pi +2(N^3-6N^2+4N-4)m^2_\pi \biggr) \nonumber \\ && \times
  A_0(m^2_\pi)
  -\frac{\Delta f_K}{f} - \frac{2\Delta
f_\pi}{f} + \frac{m_K}{2(m_K+m_\pi)}\left(\frac{\Delta
m^2_K}{m^2_K}\right) \nonumber \\ &&
 \left. +\frac{m_K+2m_\pi}{2(m_K+m_\pi)}\left(\frac{\Delta
m^2_{\pi}}{m^2_{\pi}}\right)\right] \nonumber \\
& & +\frac{2i\alpha_2}{f^3} (m_K+m_\pi)(2m_\pi-m_K)
\left[\frac{2m_K(m_K+m_\pi)}{N16 \pi^2
f^2} \right. \nonumber \\
&& -\frac{4}{N}m_K
m_\pi(m^2_K-m^2_\pi)C_0(0,q^2,q^2,m^2_{33},m^2_{33},m^2_K)\nonumber
\\ && -\frac{6}{N}m_K m_\pi B_0(q^2,
m^2_K, m^2_{33}) + 2(N-2)m_K m_\pi B_0(q^2,
m^2_K, m^2_\pi)\nonumber \\ && - 4 m^2_\pi B_0(4m^2_\pi, m^2_K, m^2_K)
-\frac{2}{N^2}(N-2)m_K m_\pi B_0(4m^2_\pi, m^2_\pi, m^2_{33}) \nonumber \\
& & -\frac{4}{N^2}m^2_\pi B_0(4m^2_\pi, m^2_{33},
m^2_{33})-\frac{2}{N^2}(N^3+2N^2-2N+2)m^2_\pi \nonumber
\\ && \times B_0(4m^2_\pi, m^2_\pi, m^2_\pi) +\frac{N
m_K}{m_\pi-m_K}A_0(m^2_K)\nonumber
\\ && +\frac{(1-N^2)m^2_K+(N^2-1)m_K m_\pi+2N^2m^2_\pi}{N(m^2_K-m^2_\pi)}
A_0(m^2_\pi)\nonumber \\ && + \frac{m_K
m_\pi(4m_K+3m_\pi)}{N(m_K+m_\pi)(2m^2_K-m^2_\pi)} A_0(m^2_{33})
 - \frac{\Delta f_K}{f} - \frac{2\Delta f_\pi}{f} \nonumber
\\ &&  + \frac{m_K(-2m_K+m_\pi)}{2(m_K+m_\pi)(2m_\pi-m_K)}\left(\frac{\Delta
m^2_K}{m^2_K}\right)
+\frac{m_\pi(m_K+4m_\pi)}{2(m_K+m_\pi)(2m_\pi-m_K)} \nonumber
\\ &&  \times \left(\frac{\Delta m^2_{\pi}}{m^2_{\pi}}\right) +\frac{32m_K
m_\pi}{f^2}\biggl(2L_1+2L_2+L_3 \biggr. \nonumber \\ &&
+\frac{m_\pi((N-4)m_K-2Nm_\pi)}{2m_K(m_K-2m_\pi)}L_4 +
\frac{m^3_K-m^2_K m_\pi-2m^3_\pi}{2m_K m_\pi(m_K-2m_\pi)}L_5
\nonumber \\ && \left. \biggl. -\frac{2m^2_K+Nm_K
m_\pi-2Nm^2_\pi}{m_K(m_K-2m_\pi)}L_6 -\frac{m^3_K-m^2_K m_\pi+m_K
m^2_\pi-2m^3_\pi}{m_K m_\pi(m_K-2m_\pi)}L_8 \biggr) \right].
\nonumber \\ \eea

\bigskip
\bigskip
\bigskip

\appendix
\section*{\bf APPENDIX F}
\setcounter{equation}{0} \setcounter{section}{1}
\renewcommand{\theequation}{F\arabic{equation}}
\bigskip

The absence of the $\alpha_2$ terms in $K\to\pi\pi$ at UK1
requires some corrections to \cite{laiho}, presented here.  Eq
(31) of \cite{laiho} should be

\begin{eqnarray}\label{20}
\langle \pi^{+} \pi^{-}|{\cal O}^{(8,1)}|K^{0}\rangle_{ct}& =& 8i
\frac{\alpha_{1}}{f^{3}}m^{2}
  + 8i\frac{m^{4}}{f^{3}}[4e^{r}_{10}+ 2e^{r}_{11}+4e^{r}_{15}
 -4e^{r}_{35}]. \nonumber \\ &&
\end{eqnarray}

\noindent In Appendix D, Eq (D6) of \cite{laiho}, the correct
equation should read

\bea \bra\pi^+\pi^- | {\cal O}^{(8,1)} | K^0\ket_{log} & = & 8i
\frac{\alpha_1}{f^3} m^2 \left[ -\frac16 m^2 \frac{1}{16\pi^2f^2}
\biggl(50\ln \frac{m^2}{\mu^2} -37\biggr) - \frac{3\Delta f}{f} +
\frac{\Delta m^2}{m^2} \right]. \nonumber
\\ &&
\eea

\end{document}